\documentclass[12pt]{article}
\usepackage[utf8]{inputenc}

\usepackage{graphicx}
\usepackage{amsmath,amssymb}
\usepackage{latexsym}
\usepackage{subfigure}
\usepackage{wrapfig}
\usepackage{color}
\usepackage{float}
\usepackage{hyperref}
\usepackage{comment}
\usepackage{soul}
\usepackage{authblk}

\newcommand{\R}{\mathbb{R}}

\newcommand{\bfa}{{\bf a}}

\newcommand{\bfd}{{\bf d}}

\newcommand{\bfm}{{\bf m}}
\newcommand{\bfn}{{\bf n}}

\newcommand{\bfu}{{\bf u}}

\newcommand{\bfx}{{\bf x}}

\newcommand{\bfE}{{\bf E}}

\newcommand{\bfH}{{\bf H}}

\newcommand{\eps}{{\varepsilon}}

\newcommand{\beq}{\begin{equation}}
\newcommand{\eeq}{\end{equation}}
\newcommand{\beqs}{\begin{eqnarray}}
\newcommand{\eeqs}{\end{eqnarray}}

\newcommand{\calE}{{\cal E}}

\usepackage[backend=biber,style=nature]{biblatex}
\title{Design of soft magnetic materials}
\author[1,*]{Ananya Renuka Balakrishna}
\author[2]{Richard D. James}
\affil[1]{\small{Aerospace and Mechanical Engineering, University of Southern California, Los Angeles, CA 90089}}
\affil[2]{\small{Aerospace and Engineering Mechanics, University of Minnesota, Minneapolis, MN 55455}}
\affil[*]{\small{Corresponding author, Office: (213) 740-8762, Email: renukaba@usc.edu}}
\date{}

\begin{document}

\maketitle
\begin{abstract}
We present a strategy for the design of ferromagnetic materials with exceptionally low magnetic hysteresis, quantified by coercivity. In this strategy, we use a micromagnetic algorithm that we have developed in previous research and which has been validated by its success in solving the ``Permalloy Problem''---the well-known difficulty of predicting the composition 78.5\% Ni of lowest coercivity in the Fe-Ni system--- and by the insight it provides into the ``Coercivity Paradox'' of W. F. Brown. Unexpectedly, the design strategy predicts that cubic materials with large saturation magnetization $m_s$ and large magnetocrystalline anisotropy constant $\kappa_1$ will have low coercivity on the order of that of Permalloy, as long as the magnetostriction constants $\lambda_{100}, \lambda_{111}$ are tuned to  special values. The explicit prediction for a cubic material with low coercivity is the dimensionless number $(c_{11}-c_{12}) \lambda_{100}^2/\kappa_1 = 81$ for $\langle 100 \rangle$ easy axes. The results would seem to have broad potential application, especially to magnetic materials of interest in energy research.
\end{abstract}

\section*{Introduction \label{sec:Introduction}}

A longstanding puzzle in materials science is understanding the origins of magnetic hysteresis in ferromagnetic materials.
Hysteresis in this domain refers to the differing behaviors obtained when a demagnetized specimen is subject to an increasing magnetic field to saturation vs.~that obtained when decreasing the field from saturation to zero. The effect is typically characterized by the final (absolute) value of the magnetic field after such a test, termed the coercivity. Informally, soft magnets have low coercivity. This paper is concerned with the prediction of coercivity from micromagnetic theory. An unexpected prediction
of our study is that coercivity can be made very small even in materials with large \textcolor{black}{magnetocrystalline anisotropy constant}, as long as the magnetostrictive constants are tuned appropriately.

Aside from basic scientific interest on the origins of hysteresis
and the traditional application to transformers, 
a strategy for the discovery of new soft magnetic materials is desirable for rapid power conversion electronics, all-electric vehicles, and wind turbines, especially in cases where induction motors are favored.
Magnetic hysteresis has also become critical to 
the adoption of proposed spintronic and storage devices, as requirements for limiting energy consumption have moved to the forefront
\cite{fert2008nobel, silveyra2018soft}.  These
requirements impact a wide range of applications,
from hand held electronic devices to storage systems
and servers at data centers.
While our analysis is aimed primarily
at bulk applications, the key idea that micromagnetic theory with magnetostriction, together with a well-chosen potent defect, can be used to predict hysteresis suggests a strategy for the lowering of coercivity also in these small-scale applications.
In fact, in certain film based devices, the likely potent defects, such as threading dislocations in 
epitaxial films, are often better 
characterized than in bulk material.

Currently, a widely accepted strategy to lower the hysteresis in cubic ferromagnetic alloys is based on changing composition so as to reduce the magnitude of the anisotropy constant $|\kappa_1|$. This has the effect of flattening the graph of
\textcolor{black}{magnetocrystalline anisotropy} energy vs.~magnetization and reducing the
penalty associated with magnetization rotation.
Intuitively, this makes sense, as it apparently makes available additional low energy pathways  of an alloy in a metastable state on the shoulder of the hysteresis loop, as
an applied field is being lowered.  A related idea is the
known strategy of tuning the composition so as to
be precisely at the point where two different symmetries coincide, again leading to a flattening of the of the \textcolor{black}{magnetocrystalline anisotropy} energy vs.~magnetization and the
lowering of hysteresis. The latter is the strategy used by Clarke and collaborators \cite{teter1990magnetostriction, clark1988magnetostriction} that led to the particular composition of Terfenol: Tb$_x$Dy$_{1-x}$Fe$_2$, $x = 0.3$. \textcolor{black}{We add that modern research on
these RFe$_2$ cubic Laves phase materials has focused
on the benefits of exploiting a nearby morphotropic phase boundary in these systems to enhance magnetostrictive response 
under small fields \cite{yang2010large,bergstrom2013morphotropic,hu2021room}.}

However, these strategies cannot be the whole story behind coercivity.  For example, in the iron-nickel system a sharp drop in coercivity occurs at the Permalloy composition, at which $\kappa_1=-161\mathrm{Jm^{-3}}$
\cite{bozorth1953permalloy}.  Tuning the \textcolor{black}{magnetocrystalline anisotropy} constant to zero in iron-nickel alloys in fact leads to an alloy composition with noticeably higher hysteresis. Similarly, in Sendust \textcolor{black}{(Fe$_{0.85}$Si$_{0.096}$Al$_{0.054}$ alloy)} the hysteresis is minimum when the \textcolor{black}{magnetocrystalline anisotropy} and magnetostriction constants are close to zero, and not precisely at $\kappa_1 = 0$ \cite{takahashi1987magnetocrystalline}. Finally, there are quite a few isolated examples of alloys which have very large \textcolor{black}{magnetocrystalline anisotropy} constants  but low coercivity: an example is the uniaxial phase of Ni$_{51.3}$Mn$_{24.0}$Ga$_{24.7}$ having a uniaxial \textcolor{black}{magnetocrystalline anisotropy} constant of $2.45 \times 10^5$ Jm$^{-3}$ and coercivity less than $1$ kAm$^{-1}$ in single crystals \cite{tickle1999magnetic}. \textcolor{black}{Another example is the Galfenol alloys ($\mathrm{Fe_{1-x}Ga_{x}}$ for $0.13 \leq x \leq 0.24$) that have small magnetic hysteresis despite their very large \textcolor{black}{magnetocrystalline anisotropy} constants \cite{atulasimha2011review, clark2002magnetostrictive}.} These examples suggest that the \textcolor{black}{magnetocrystalline anisotropy} constant is the not the only factor that governs hysteresis in magnetic alloys.

The precise role of magnetostriction constants on magnetic hysteresis is not well understood for two  reasons: (a) Prior research has typically ignored the contribution of magnetoelastic interactions on hysteresis, justified by the small values of the magnetostriction constants, and (b) mathematical methods, such as the linear stability analysis,  overestimate the coercivity of bulk alloys---by over three orders of magnitude in some cases---as compared to measured experimental values \cite{brown1963micromagnetics, aharoni1958magnetization}. \textcolor{black}{The latter is
referred to as the ``Coercivity Paradox'' \cite{brown1963micromagnetics, brown1962magnetostatic}}.  These factors limit our understanding of how the balance between fundamental material constants---such as \textcolor{black}{magnetocrystalline anisotropy} and magnetostriction constants---affects magnetic coercivity.

\begin{figure}
\begin{centering}
\textcolor{black}{\includegraphics[width=0.7\textwidth]{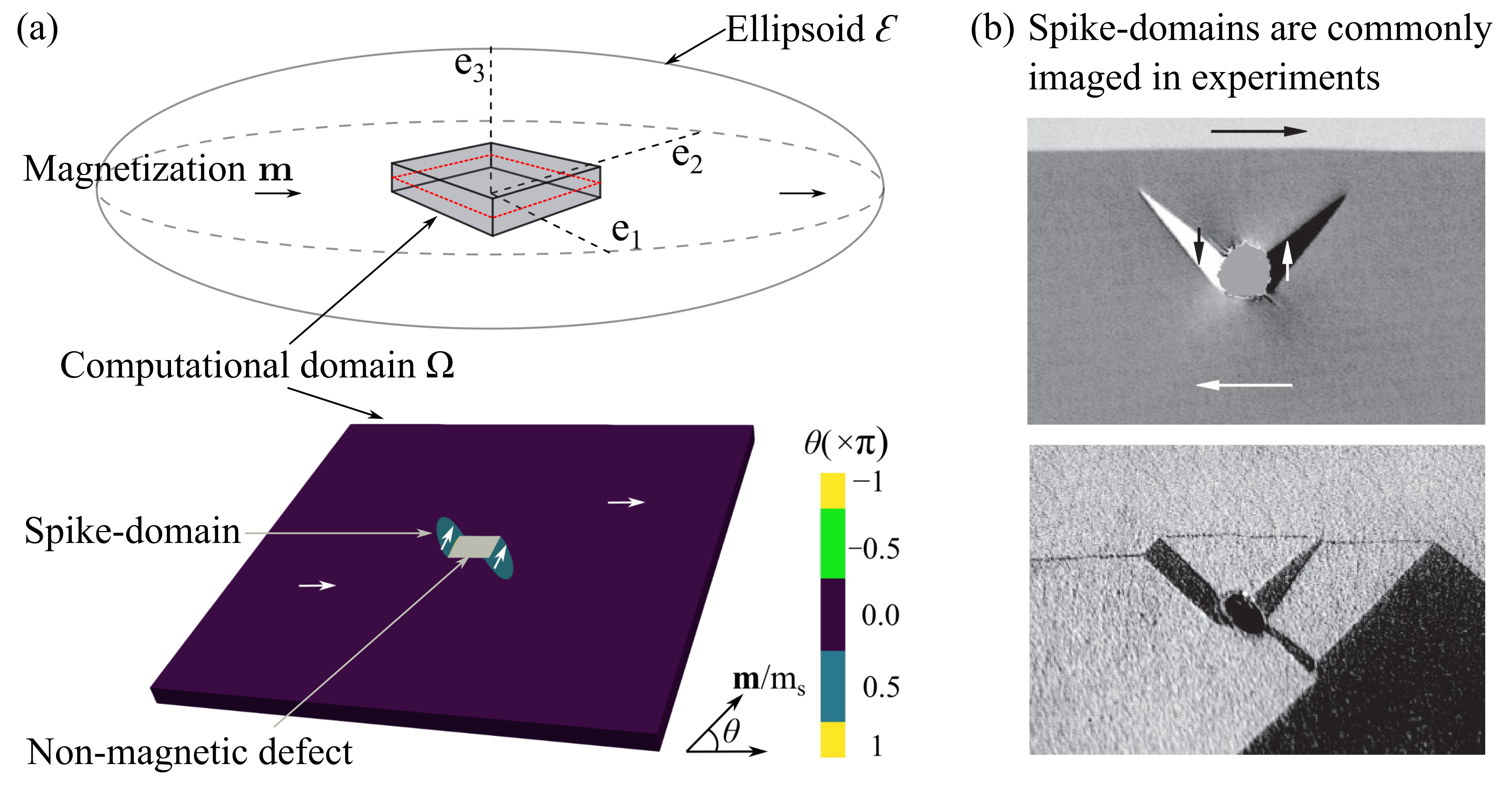}}
\par\end{centering}
\caption{We model a 3D computational domain with a spike-domain microstructure. (a) The 3D computational domain is $\Omega$ embedded inside an oblate ellipsoid $\calE$, which is several times larger in size than $\Omega$. A spike-domain microstructure forms around a non-magnetic defect that grows under an applied field. (b) Spike-domain microstructures are commonly found in magnetic materials around defects such as cavities and pores in Fe-film and Fe-Si crystal, respectively. \cite{schafer2020tomography,hubert2008magnetic}. \textcolor{black}{(Reprinted subfigure (b) with permission from Ref. \cite{schafer2020tomography}. Copyright (2020) by the American Physical Society)}} \label{Fig:1}
\end{figure}

In our recent work, we developed a computational tool based on micromagnetics, including magnetoelastic terms, that is adapted to the prediction of  coercivity in bulk magnetic alloys \cite{balakrishna2021tool, renuka2021solution}. \textcolor{black}{Micromagnetics, since its first postulation in 1963 by W.F.Brown Jr. \cite{brown1963micromagnetics}, has been applied to a wide range of problems in ferromagnets and forms the basis to several computational frameworks \cite{wang2001gauss,zhang2005phase,wang2013real}. Our coercivity tool is based on the micromagnetics theory \cite{balakrishna2021tool}, however differs from earlier works in the following ways: A key feature of this tool is that it uses a large but highly localized disturbance---in the form of a N\'eel-type spike domain---to predict magnetic coercivity. N\'eel spike-like domains are frequently observed to form around defects in various magnetic alloys (See \cite{hubert2008magnetic, otto2010domain,otto2010concertina,doring2014reduced,cinti2016interpolation} for  examples) and grow under an external field. In the absence of this spike-domain, the second variation of the micromagnetic energy misses the energy barrier for magnetization reversal \cite{brown1966magnetoelastic, brown1963micromagnetics}. Instead, we model a large localized disturbance (i.e., compute a non-linear stability analysis) to predict coercivity on the shoulder of the hysteresis curve, see Fig.~\ref{Fig:1}. Other features of our coercivity tool are that we account for magnetoelastic interactions, however small, in estimating coercivity of bulk magnetic alloys; and we use the ellipsoid and reciprocal theorems to accelerate our coercivity calculations, see \cite{balakrishna2021tool}. Our computations are fully three-dimensional with the magnetization evolving from (010) to (100) in all directions as one leaves the spike domain.}

\textcolor{black}{To arrive at the N\'eel spike as a  reasonable description of a potent defect, we include in \cite{balakrishna2021tool} a study of alternative nuclei, including rotated spikes and multiple spikes.  These spike domains serve as nuclei that grow during magnetization reversal. In this previous study \cite{balakrishna2021tool}, we investigated the role of defect geometry, defect orientation, and defect number on magnetic coercivity, and we found that these features did not significantly affect coercivity when compared to fundamental material constants. Mathematically, they can be interpreted as localized disturbances that, under the
right combination of material constants and suitable
applied field, are able to surmount an energy barrier \cite{pilet2006relation, knupfer2013nucleation, zhang2009energy, zwicknagl2014microstructures}} Small, smooth disturbances of a homogeneous state are not
able to surmount this barrier at 
realistic applied fields due to
the dominance of domain wall energy
at small scales, and, in our view,  this is the essence
of the Coercivity Paradox
\cite{balakrishna2021tool,renuka2021solution}. 
Mathematically, the situation is that the study of
the second variation of the micromagnetic energy misses the important energy barrier,  while it is 
captured as a large amplitude, but localized, disturbance.

Our viewpoint is  consistent with the thesis of Pilet \cite{pilet2006relation} who examines the microscopic state on the shoulder of the hysteresis loop (far away from where coercivity is measured) using magnetic force microscopy, and finds good correlation with the presence of large localized disturbances. Another feature of our tool is that we account for the magnetoelastic interactions---however small---in estimating coercivity. Our results are in a form that is amenable to alloy development, as in related searches for low hysteresis phase transformations \cite{cui2006combinatorial}.
 
An alternative highly effective route to low hysteresis in magnetic materials is the synthesis of nanocrystalline material \cite{thomas2020nanocrystallites}.  For example, the powder synthesis and rapid solidification techniques have resulted in nanocrystalline and amorphous magnetic alloys that offer low hysteresis and enhanced permeability \cite{mchenry1999amorphous,silveyra2018soft,soldatov2020inverted}. Here we do not specifically analyze this situation, but 
we think this should be possible in the nanocrystalline case, though challenging
due to the many grains that would likely have to be considered.

The aim of the present work is to identify combinations of material constants, including saturation magnetization ($m_s$), \textcolor{black}{magnetocrystalline anisotropy} constant ($\kappa_1$), elastic moduli ($c_{11}, c_{12}, c_{44}$), and magnetostriction
constants ($\lambda_{100}, \lambda_{111}$) that give
lowest coercivity in a cubic material.  \textcolor{black}{We do not
vary the exchange constant, which in practice does not vary much: we
also give arguments why, when combined with the
demagnetization energy, it should matter little.} We find the striking result that magnetostriction plays a critical role, and more importantly, coercivities on
the order of those found in very soft materials such
as Permalloy are predicted to be possible in materials
with large \textcolor{black}{magnetocrystalline anisotropy} constant, as long as the
magnetostriction constants are tuned to special values.
 We find that \textcolor{black}{magnetocrystalline anisotropy}
and magnetostriction constants play a particularly 
important role, but neither has to be small.

The simulations are carried out using our newly developed coercivity tool.  Details on our micromagnetic algorithm are described in Methods. Specifically, we apply this tool in two studies: In Study 1, we test the hypothesis that the tuning of  magnetostriction constants, in addition to the \textcolor{black}{magnetocrystalline anisotropy} constant, is necessary to reduce coercivity in magnetic alloys. Here we study how combinations of $\kappa_1$ and $\lambda_{100}$ affect coercivity, and ignore the contribution from $\lambda_{111}=0$. In total, we compute coercivity values from $N = 2,163$ independent simulations. In Study 2, we test our hypothesis that there exists a specific combination of material constants---$\kappa_1, \lambda_{100}, \lambda_{111}$---at which magnetic coercivity is the lowest. Here, we compute coercivity by systematically varying the \textcolor{black}{magnetocrystalline anisotropy} and the magnetostriction constants, in the ranges $ 0 \leq \kappa_1 \leq 2000\mathrm{Jm^{-3}}$, $ -2000\mu\epsilon \leq \lambda_{100} \leq 2000\mu\epsilon $, and $ 0 \leq \lambda_{111} \leq 600\mu\epsilon $ respectively ($N = 605$). Overall, our computations from over 2500 independent simulations show that the lowest coercivity is attained when the dimensionless number  $\frac{(c_{11}-c_{12})\lambda_{100}^2}{2 \kappa_1} \approx 81$. Here, $c_{11}, c_{12}$ are the elastic stiffness constants of the soft magnet assuming a linear cubic relation. To our knowledge, this discovery of the balance between material constants at which magnetic hysteresis is small has not been proposed before.  A theoretical analysis supporting
the importance of this dimensionless number is given
below.  This analysis further supports the idea
that, in other situations,  a second dimensionless 
number $\frac{c_{12}\lambda_{100}^2}{2 \kappa_1}$
may become important, and tuning the stiffness
constants $c_{11}$ and  $c_{12}$ so that both of
these dimensionless constants have particular values
may be desirable \cite{ahani2021}.

\section*{Results\label{sec:Results}}

\subsection*{Computation of coercivity}
\begin{figure}
\includegraphics[width=0.95\textwidth]{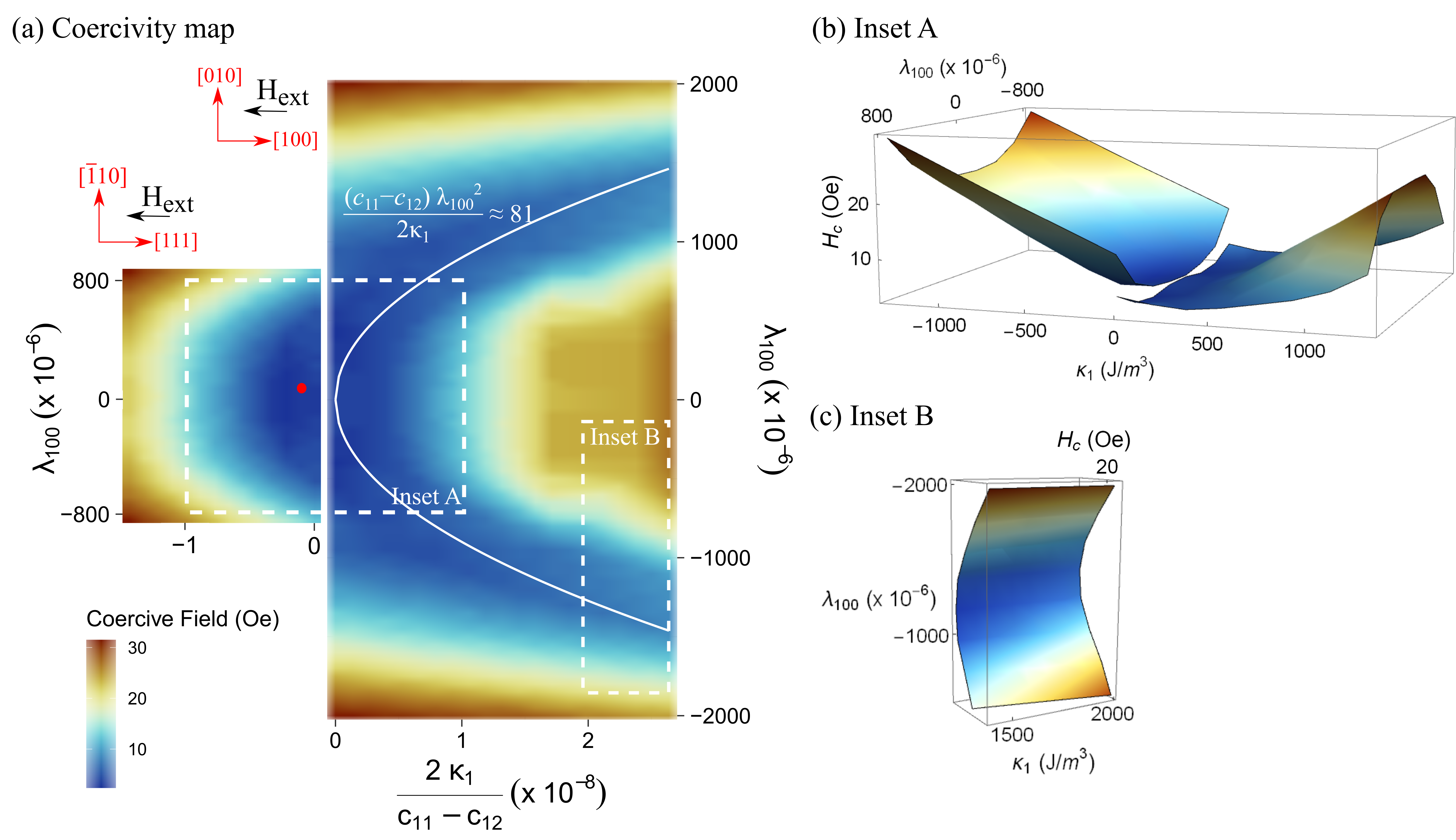}
\caption{The coercivity map as a function of the \textcolor{black}{magnetocrystalline anisotropy} $\kappa_{1}$ and the magnetostriction constant $\lambda_{100}$. (a) We carried out a total of $N=2163$ computations with the magnetostriction constant $\lambda_{111}$ set to zero. \textcolor{black}{We find that minimum coercivity is achieved when $\frac{(c_{11}-c_{12})\lambda_{100}^2}{2 \kappa_1} \approx 81$.} For comparison, the solid red-dot corresponds to the coercivity of the permalloy composition. \textcolor{black}{Note, the parabolic relationship is a best fit polynomial to the spread of coercivity data from our calculations. Alloys, such as the permalloy, lie close to the vertex of this parabola and we interpret this as the shortest distance to the parabola with a dimensionless constant of 81.} The axes (in red) indicates the crystallographic directions along which coercivity values were measured for alloys with $\kappa_1 >0$ and $\kappa_1 < 0$, respectively. The 3D surface contours of the coercivity values in (b) `inset A' and (c) `inset B' are shown. These plots show coercivity wells (regions of minimum coercivity) as $\kappa_{1}\to 0$ and $\kappa_{1}>>0$, respectively. Note, in sub-figure (b), the 3D surface contour is discontinuous along $\kappa_1 = 0$---we attribute this discontinuity to the transformation of easy axes from [100] to [111] crystallographic direction.\label{Fig:2}}
\end{figure}

In Study 1, we test our hypothesis that the magnetostriction constants, in addition to the \textcolor{black}{magnetocrystalline anisotropy} constant, is necessary to reduce coercivity in magnetic alloys. 
Fig.~\ref{Fig:2}(a) shows a heat map of coercivity values as a function of the \textcolor{black}{magnetocrystalline anisotropy} constant $\kappa_1$ and the magnetostriction constant $\lambda_{100}$. \textcolor{black}{A key feature of this plot is that the coercivity is minimum along a curve described by   $\frac{(c_{11}-c_{12})\lambda_{100}^2}{2 \kappa_1} \approx 81$.} As expected the coercivity is small when $\kappa_1 \to 0$ and the magnetostriction constant is small $\lambda_{100} \to 0$, see Inset A. However, surprisingly, we find that the coercivity value is also small for non-zero \textcolor{black}{magnetocrystalline anisotropy} constants $\kappa_1 >> 0$ with suitable combinations of the magnetostriction constant, see Inset B.  Note the
rather large \textcolor{black}{magnetocrystalline anisotropy} constants being considered, well outside the range associated to normal soft magnetism.  \textcolor{black}{Furthermore, the minimum coercivity valleys are symmetric about the $\lambda_{100} = 0$ axis. This symmetric response arises from the even terms $\lambda_{100}^2$ in the free energy expression.} \textcolor{black}{We observe a similar lowering of coercivity in magnetic alloys with $\kappa_1 < 0$ at suitable values of $\lambda_{111}$ magnetostriction constant (see Supplementary Figure 2).}

Fig. \ref{Fig:2}(b-c) are 3D surface plots of the inset regions `A' and `B' respectively. The ``wells" in these plots correspond to combinations of material constants at which coercivity is minimum. Fig. \ref{Fig:2}(b) is a 3D plot of the inset region `A'---here, we note a discontinuity or a jump in coercivity values at $\kappa_1 = 0$. This discontinuity is because of the change in easy axes for magnetic alloys with $\kappa_1>0$ and $\kappa_1<0$. For example, we compute coercivities along $[100]$ and $[111]$ crystallographic directions for magnetic alloys with $\kappa_1>0$ and $\kappa_1<0$, respectively. These alloys have different values of magnetostriction constants along their easy axes. We believe that
the computed rapid change of coercivity at $\kappa_1 = 0$ is real, and arises from the anisotropy of these magnetostriction constants. Fig. \ref{Fig:2}(c) shows the 3D surface plot of the inset region B. Here, coercivities comparable to that of the permalloy composition are achieved at large \textcolor{black}{magnetocrystalline anisotropy} values $\kappa_1 \approx 1700\mathrm{Jm^{-3}}$ with suitable combinations of the magnetostriction constant $\lambda_{100} \approx 1200\mu\epsilon$.

Overall, the findings from Study 1 contradict the general understanding of hysteresis in magnetism, i.e., the \textcolor{black}{magnetocrystalline anisotropy} constant needs to be near zero for small hysteresis. Our findings show that magnetic hysteresis (or coercivity value) is small not only when $\kappa_1 \to 0$ but also when $\kappa_1 >> 0$, given suitable values of the magnetostriction constants. Fig.~\ref{Fig:2} shows that the magnetostriction constant, $\lambda_{100}$, in addition to the \textcolor{black}{magnetocrystalline anisotropy} constant $\kappa_1$ plays an important role in reducing coercivity in magnetic alloys. 

\begin{figure}
\includegraphics[width=0.95\textwidth]{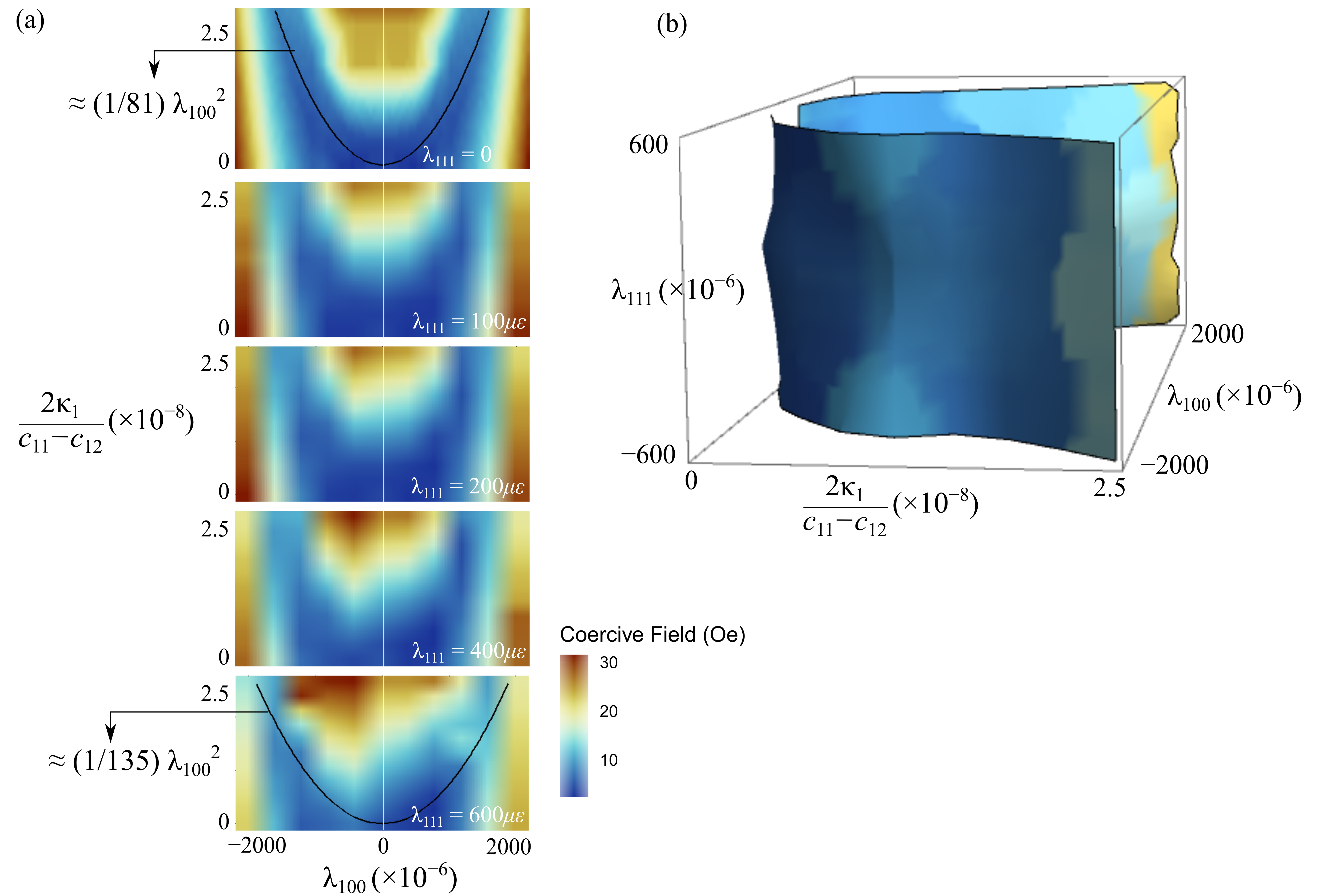}
\caption{The coercivity map as a function $\kappa_{1}$, $\lambda_{100}$ and $\lambda_{111}$. (a) \textcolor{black}{The minimum coercivity is achieved along a parabolic relation,   $\frac{(c_{11}-c_{12})\lambda_{100}^2}{2 \kappa_1} \approx p$ . Here, the value of coefficient $p$ depends on the value of the second magnetostriction constant $\lambda_{111}$.} For example, as $\lambda_{111}$ increases and the coercivity well loses symmetry across the $\lambda_{100}=0$ axes. (b) The 3D surface plot identifies a combination of material constants, namely $\kappa_1$, $\lambda_{100}$ and $\lambda_{111}$, at which the coercivity is small.\label{Fig:3}}
\end{figure}

Thus far, we investigated coercivity values by setting one of the magnetostriction constants $\lambda_{111}$ to be zero. In Study 2 we test the hypothesis that there exists a specific combination of material constants, namely $\kappa_1, \lambda_{100}, \lambda_{111}$, at which magnetic coercivity is the lowest. We compute coercivities at every combination of $\kappa_1,\lambda_{100}$ and $\lambda_{111}$ in the parameter space $0 \leq \kappa_1 \leq 2000\mathrm{Jm^{-3}}$ and $-2000\mu\epsilon \leq \lambda_{100} \leq 2000\mu\epsilon$ and $0 \leq \lambda_{111} \leq 600\mu\epsilon$.

Fig. \ref{Fig:3}(a) shows the coercivity map as a function of $\kappa_1$ and $\lambda_{100}$ for increasing values of $\lambda_{111}$. A key feature here is that the minimum coercivity relationship $\kappa_1 \propto \lambda_{100}^2$ is unique at each value of the $\lambda_{111}$ magnetostriction constant. For example, the minimum coercivity valleys gradually widen and become asymmetric about $\lambda_{100}=0$, with increasing value of the $\lambda_{111}$ magnetostriction constant. We identify combinations of material constants at which minimum coercivity is achieved, and then plot a 3D surface through these points in the $\kappa_1-\lambda_{100}-\lambda_{111}$ parameter space, see  Fig. \ref{Fig:3}(b). This 3D plot represents a surface through the material parameter space on which coercivity is small.

\subsection*{Theoretical analysis \label{sec:Theory}}

The results of our studies above can be understood in the following way.  We begin  from the free energy 
that we have used in the simulations of
micromagnetics in dimensional form \cite{balakrishna2021tool},
\beqs
\lefteqn{\int_{\Omega}\{\nabla\mathbf{m}\cdot\mathrm{A\nabla\mathbf{m}+\kappa_{1}(\mathrm{m_{1}^{2}m_{2}^{2}}+
\mathrm{m_{2}^{2}m_{3}^{2}}+\mathrm{m_{3}^{2}m_{1}^{2}})}} \nonumber \\
& & +\  \frac{1}{2}[\mathbf{E}-\mathbf{E_{0}\mathrm{(\mathbf{m})}}]\cdot\mathbb{C}[\mathbf{E}-\mathbf{E_{0}\mathrm{(\mathbf{m})}}]-\mu_{0}m_s\mathbf{H_{\mathrm{ext}}\cdot m}\}\mathrm{d\mathbf{x}}
  +\frac{1}{2}\int_{\mathbb{R}^{3}}\mu_{0}\left|\mathbf{H_{d}}\right|^{2}\mathrm{d\mathbf{x}}.  \label{eq:FreeEnergy1}
\eeqs
where $\bfE = \frac{1}{2}(\nabla \bfu + (\nabla \bfu)^T)$
and the energy is to be minimized, or minimized locally,
over the pair of functions $\bfu, \bfm \in H^1(\Omega)$.
Here, $\bfm$ (with components on the cubic axes $\mathrm{m_1, m_2, m_3}$) has been previously non-dimensionalized so $\bfm \cdot \bfm = 1$,
and the demagnetization field satisfies the magnetostatic
equations $\bfH_{\bfd} = - \nabla \zeta_{\bfm}, 
\nabla \cdot (- \nabla \zeta_{\bfm} + m_s \bfm ) =0$
on all of space, so $\bfH_{\bfd}$ has the dimensions of $m_s$, as does 
$\bfH_{\rm ext}$. Please note that $\bfm$ is extended to $\R^3$ by making it vanish outside $\Omega$. Typical accepted values from a large compositional space  
appropriate to Fig.~\ref{Fig:3}(a) including most 
of the Fe-Ni system are
\beqs
&{\rm A} \sim 10^{-11} \mathrm{Jm^{-1}},\ \ \kappa_1 \sim 0-6 \times 10^3 \mathrm{Nm^{-2}}, \ \ \lambda_{100} \sim 0-2000 \times 10^{-6},\ \ \lambda_{111}
\sim 0-600 \times 10^{-6}, & \nonumber \\
& m_s \sim 10^6 \mathrm{Am^{-1}}, \mu_0 \sim 1.3 \times 10^{-6}\mathrm{NA^{-2}},\ \  c_{11}, c_{12}, c_{44}
\sim 10-24 \times 10^{10} \mathrm{Nm^{-2}}&
\eeqs
A  nondimensional form is obtained by dividing
the micromagnetic energy by 
$\mu_0 m_s^2 \sim  10^{6} \mathrm{Nm^{-2}}$, changing variables $\bfx'= \frac{1}{\ell}\, \bfx \in \Omega'$, where $\ell$ is a typical length scale and $\bfx'$ is dimensionless. This gives a typical nondimensional values of the micromagnetic coefficients
\beqs
&\frac{A}{\ell^2 \mu_0 m_s^2} \sim 10^{-17},\ \ \frac{\kappa_1}{\mu_0 m_s^2} \sim 10^{-3}, \ \ \frac{(c_{11} - c_{12} \ {\rm or}\ c_{44})( \lambda_{100}^2 \ {\rm or}\ \lambda_{111}^2)}{\mu_0 m_s^2}
\sim 10^{-5} - 10^{-1},  & \nonumber \\
&\frac{1}{m_s} H_{\rm ext} \sim 10^{-3},\ \  \frac{ 1}{m_s^2}H_{\rm d}^2
\sim 1&  \label{nondim}
\eeqs
The range of magnetostrictive coefficients is consistent with
Fig.~\ref{Fig:3}(a), and for material constants with a
size range, we choose a typical intermediate value. A caveat with these numbers is the observation made by Brown \cite{brown1966magnetoelastic} (discussed also
in \cite{hubert2008magnetic}) that the formal
linearization of geometrically nonlinear micromagnetics
that gives Eq.~\ref{eq:FreeEnergy1} implies a possible
$\bfm$ dependence of the elastic moduli $c_{11}, c_{12},
c_{44}$.  This dependence is usually neglected, as is done here.  Note from Eq.~\ref{nondim} the wide ranging values of magnetostrictive energy and its relative importance generally.

It is seen from the nondimensionalized coefficients
Eq.~\ref{nondim} that magnetostriction and demagnetization
energy are dominant, but it is important to observe
that each can be made negligible
by suitable magnetization distributions.  The magnetostrictive energy can be made to vanish
by choosing a magnetization that satisfies curl$({\rm curl}\, \bfE_0(\bfm))^T = 0$, that is, $\bfE_0(\bfm(\bfx)) = \frac{1}{2}(\nabla \bfu + (\nabla \bfu)^T)$ is the symmetric part of a gradient,
while the demagnetization energy vanishes on divergence-free magnetizations. Additionally, we note that the exchange energy is one of the smallest energy contributions and its contribution further decreases at increasing length scales. Although the exchange constant could be affected by temperature, composition, and presence of defects, its order of magnitude $A \approx 10^{-11}$Jm$^{-1}$ does not significantly affect large scale micromagnetic simulations. Consequently, we do not vary the exchange constant in our calculations.

\begin{figure}
\begin{centering}
\includegraphics[width=0.75\textwidth]{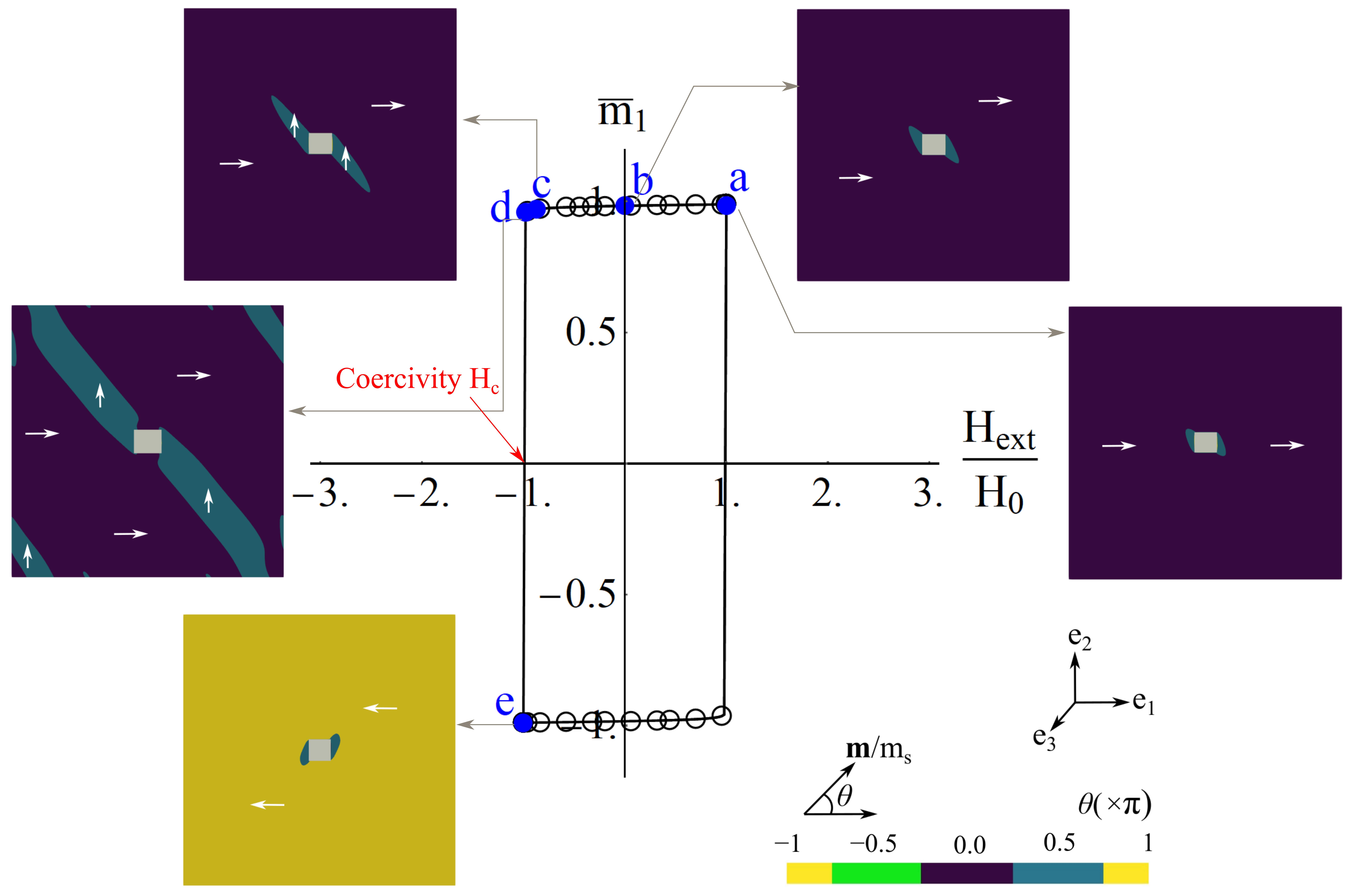}
\par\end{centering}
\caption{Growth of a spike-domain microstructure during a typical magnetization reversal. A spike-domain forms around a non-magnetic inclusion in the computation domain. This spike grows gradually on decreasing the external field (a-c). At the coercivity value, the spike-domain grows abruptly resulting in magnetization reversal (d-e). The hysteresis loop plots the average magnetization in the computation domain $\overline{\mathrm{m}}_1$ as a function of the applied field $\mathrm{H_{ext}}$. \label{Fig:4} 
\vspace{-5mm}
}
\end{figure}

We first explain from a theoretical viewpoint why the N\'eel spike attached to a defect of the type we have chosen is a particularly potent perturbation. A key
observation comes from symmetry and holds also in the
more general case of a geometrically nonlinear 
magnetoelastic free energy (see Section 6 of
\cite{james2000martensitic}).  The observation is that 
domain walls involving a jump in the magnetization
$\bfm^+ - \bfm^- \ne 0$, where $\bfm^+$ and $\bfm^-$
minimize the \textcolor{black}{magnetocrystalline anisotropy} energy density and satisfy
the divergence free condition $(\bfm^+ - \bfm^-)\cdot \bfn =0$ at an interface with normal $\bfn$, have the property
that they give rise to strains that are perfectly mechanically compatible in the sense that
$\bfE_0(\bfm^+) - \bfE_0(\bfm^-) = \frac{1}{2}(\bfa \otimes \bfn + \bfn \otimes \bfa)$ for some vector $\bfa$.
The latter is the jump condition implying the existence
of a continuous displacement across the interface.
These conditions hold also for typical domain wall models
with remote boundary conditions given by $(\bfm^+, \bfE_0(\bfm^+))$ and $(\bfm^-, \bfE_0(\bfm^-))$.
This argument applies not only to materials with cubic
symmetry but also to many lower symmetry systems (see 
\cite{james2000martensitic} for the precise statement of this
result). \textcolor{black}{.}

To understand the relation between this symmetry argument
and the N\'eel spike, we substitute the form of $\mathbb C$ and $\bfE_0(\bfm)$ for cubic materials into Eq.~\ref{eq:FreeEnergy1}.  The latter is,
\beq
\bfE_0(\bfm) = \frac{3}{2}\left(\begin{array}{ccc}
\lambda_{100}({\rm m_1^2} - \frac{1}{3}) & \lambda_{111}{\rm m_1 m_2} & \lambda_{111}{\rm m_1 m_3} \\
\lambda_{111}{\rm m_1 m_2} & \lambda_{100}({\rm m_2^2} - \frac{1}{3}) & \lambda_{111}{\rm m_2 m_3} \\
\lambda_{111}{\rm m_1 m_3} & \lambda_{111}{\rm m_2 m_3} & \lambda_{100}({\rm m_3^2} - \frac{1}{3}) \end{array} \right)
\eeq
in the orthonormal cubic basis.
We consider only the case $\kappa_1>0$ 
corresponding to $\langle 100 \rangle$ 
easy axes, for
which we have the most data above.  
(The case $\kappa_1 < 0$ corresponding to
$\langle 111 \rangle$ easy axes is handled
similarly.) 
Without loss of generality we also divide the whole energy by the dimensionless number $\kappa_1/\mu_0 m_s^2$. In addition, we define a new displacement $\hat{\bfu}(\bfx)$ by
${\bfu}(\bfx) = \lambda_{100} \hat{\bfu}(\bfx)$
with corresponding strain tensor $\bfE = \lambda_{100} \hat{\bfE}$.
Minimization of energy using $\bfu$
is equivalent to the same using $\hat{\bfu}$, and this equivalence applies
also  to local minimization or the relative height of an energy barrier. 
These changes give the
explicit nondimensional form of the free energy,
\begin{align}
& \int_{\Omega'}\mathrm{\frac{A}{\ell^2 \kappa_1}\rm{m_{i,j}m_{i,j}}+ (\mathit{\mathrm{{m}_{1}^{2}{m}_{2}^{2}}+\mathrm{{m}_{2}^{2}{m}_{3}^{2}}+\mathrm{{m}_{3}^{2} {m}_{1}^{2}}})}
\nonumber \\
 & + \frac{2c_{44} \textcolor{black}{\lambda_{100}^2}}{\kappa_1}\left[\left(\hat{\epsilon}_{12}-\frac{3}{2}\frac{\lambda_{111}}{\lambda_{100}}{\rm m_{1}m_{2}}\right)^{2}+\left(\hat{\epsilon}_{13}-\frac{3}{2}\frac{\lambda_{111}}{\lambda_{100}}{\rm m_{1}m_{3}}\right)^{2}+\left(\hat{\epsilon}_{23}-\frac{3}{2}\frac{\lambda_{111}}{\lambda_{100}}{\rm m_{2}m_{3}}\right)^{2}\right]
 \nonumber \\
 & +\left(\frac{(c_{11}-c_{12})\lambda_{100}^2}{2 \kappa_1}\right)\left[\left(\hat{\epsilon}_{11}-\frac{3}{2}\left({\rm m}_{1}^{2}-\frac{1}{3}\right)\right)^{2}+\left(\hat{\epsilon}_{22}-\frac{3}{2}\left({\rm m}_{2}^{2}-\frac{1}{3}\right)\right)^{2}+\left(\hat{\epsilon}_{33}-\frac{3}{2}\left({\rm m}_{3}^{2}-\frac{1}{3}\right)\right)^{2}\right]\nonumber \\
 & +\frac{c_{12}\lambda_{100}^2}{2 \kappa_1}
 \bigg( \hat{\epsilon}_{11} + \hat{\epsilon}_{22} + \hat{\epsilon}_{33} \bigg)^2
 \nonumber \\
  & - \frac{\mu_0 m_s}{\kappa_1}(H_{\mathrm{e}1}{\rm m}_{1}+H_{\mathrm{e}2}{\rm m}_{2}+H_{\mathrm{e}3}{\rm m}_{3}) \nonumber \\
 & -\frac{\mu_0 m_s}{2 \kappa_1}(\mathbf{\mathit{H}_{\mathrm{d1}}{{\rm m}_{\mathrm{1}}}+\mathit{H}_{\mathrm{d2}}{{\rm m}_{\mathrm{2}}}+\mathit{H}_{\mathrm{d3}}{{\rm m}_{\mathrm{3}}}}) \, \mathrm{d\bfx},\label{eq:EnergyIndicialNotation}
\end{align}
where we have used the magnetostatic equation to include
the demagnetization term in the integrand.  Also,
$\hat{\epsilon}_{ij} = \frac{1}{2}( \hat{u}_{i,j} + \hat{u}_{j,i})$ and for simplicity we have dropped the prime on $\bfx$.
However, $\Omega'$ remains dimensionless.

\begin{figure}
    \centering
    \includegraphics[width=0.8\textwidth]{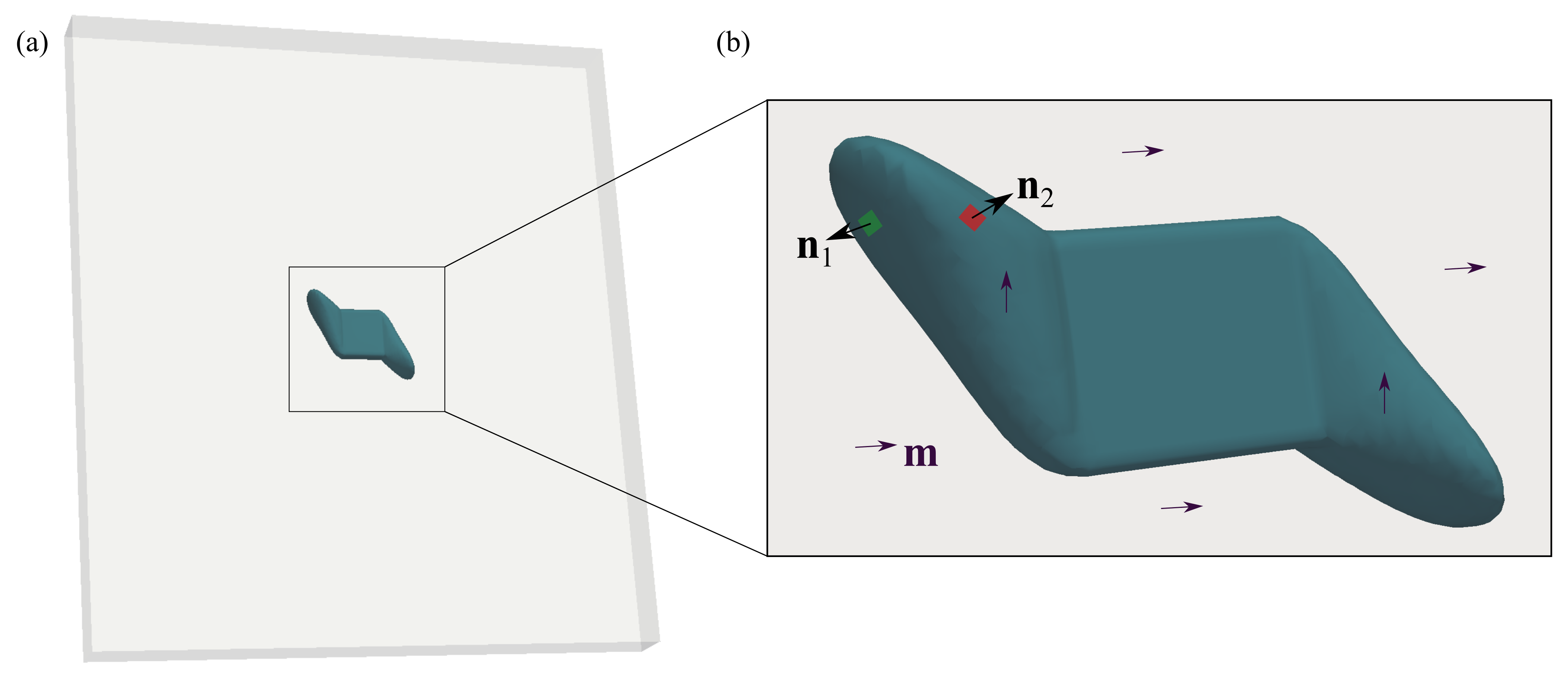}
    \caption{\textcolor{black}{A 3D visualization of the needle domain. (a) This needle domain forms around an embedded defect in our micromagnetic computations. (b) The inset figure shows the needle domain with distinct surfaces (marked by surface normal vectors $\mathbf{n}_1, \mathbf{n}_2$). The surface $\mathbf{n}_1$ approximately satisfies the strain compatibility condition, however, the surface $\mathbf{n}_2$ is not compatible and contributes to non-zero elastic energy in the system. The magnetization is denoted by $\mathbf{m}$.}}
    \label{Fig:5}
\end{figure}

\begin{figure}
\centering
\includegraphics[width=0.8\textwidth]{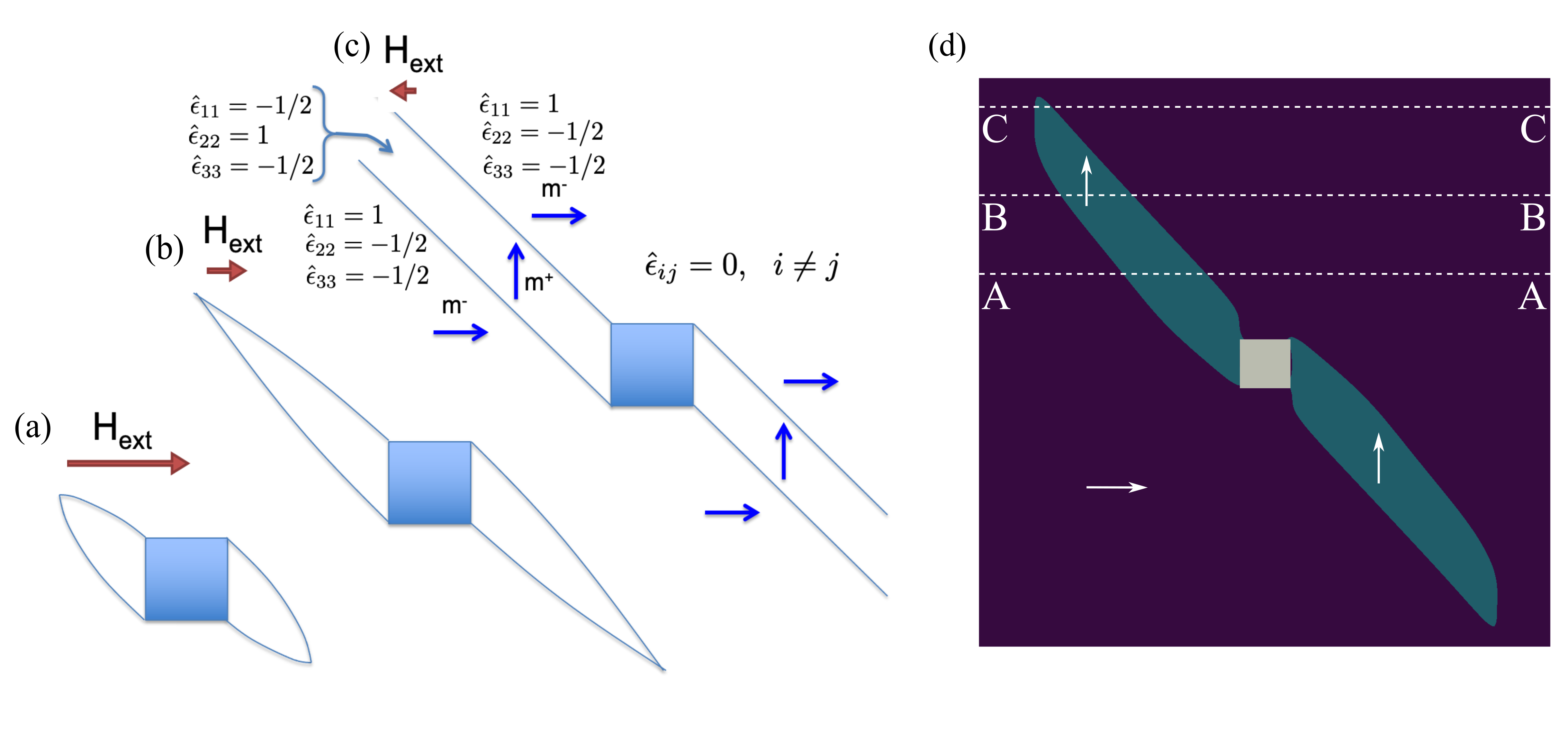}
\caption{Schematic of the metastability of the N\'eel spikes.  (a) With large applied field the spikes are collapsed onto the nonmagnetic defect, stabilized by (diffuse) domain wall and demagnetization energy.  (b) As the applied field is decreased the spikes grow, so as to decrease magnetostrictive and demagnetization energies. (c) The associated energy barrier is breached at small negative applied fields, and the spikes run to the boundary.  The state (c) is transient, and precedes the full switch to the lower energy $\langle \bar{1}00 \rangle$ magnetization; the final state is like (a) except the spike domains are pointing NE and SW (see Fig.~\ref{Fig:4}e). (d) \textcolor{black}{An example simulation of the spike domain at its transient domain state during magnetization reversal. Labels `AA', `BB', `CC' mark lengths across which we examine the strain distribution in Fig.~\ref{Fig:7}.}\label{Fig:6}}
\end{figure}

Now we make an observation about the structure of Eq.~\ref{eq:EnergyIndicialNotation} relating to the
symmetry argument given above.  With $\kappa_1>0$ as assumed,
the two magnetizations $\bfm^- = (1,0,0)$ and 
$\bfm^+ = (0,1,0)$ satisfy $(\bfm^+ - \bfm^-) \cdot \bfn = 0$, where $\bfn = \frac{1}{\sqrt{2}}(1,1,0)$.  Therefore, by the symmetry argument described above,
$\bfE_0(\bfm^+) - \bfE_0(\bfm^-) = \frac{1}{2}(\bfa \otimes \bfn + \bfn \otimes \bfa)$ and it is indeed verified that
that this holds with $\bfa = \frac{3\sqrt{2} \lambda_{100}}{2} (-1,1,0)$.
The values of the strains are given in Fig.~\ref{Fig:6}(c).
Importantly, these choices of $\bfm^+, \bfE_0(\bfm^+)$ and $\bfm^-, \bfE_0(\bfm^-)$ 
make the \textcolor{black}{magnetocrystalline anisotropy} energy and all three magnetostrictive terms in Eq.~\ref{eq:EnergyIndicialNotation} vanish,
and also give a locally divergence-free
magnetization.   Recalling that the simulations are 3D \textcolor{black}{(see Fig.~\ref{Fig:5})},
note that Fig.~\ref{Fig:6}(c) can be confined to a slab between two surfaces parallel to the plane of the page and, if 
$\bfm (\bfx) = \bfm^-$ is assigned outside the slab, then
these surfaces are also pole-free.  These facts support the potency (i.e., low energy barrier) provided by the N\'eel spikes.

\textcolor{black}{At the transient state, the magnetic poles on the tips of the spike domains are far apart and the geometric features of the spike-domain have evolved to compatible interfaces. Similar symmetry arguments can be made for transient state microstructures with multiple N\'eel spikes and/or other defect geometries. In these cases the magnetic material constants dominate the energy barrier, and the defect geometries appear to play a negligible role.}

Thus we arrive at the following scenario typical of nucleation. When the applied field is large the \textcolor{black}{magnetocrystalline anisotropy} energy of a large spike is disfavored, and spike is small and collapsed near the defect, due to the dominating influence of (diffuse) domain wall energy at small scales.  As the applied field is decreased, the \textcolor{black}{magnetocrystalline anisotropy} and
demagnetization  energies favor the
growth of the spike.  A local energy maximum is reached,
beyond which an energy decreasing path is possible, leading
 to complete reversal of the magnetization.

\begin{figure}
\centering
\includegraphics[width=\textwidth]{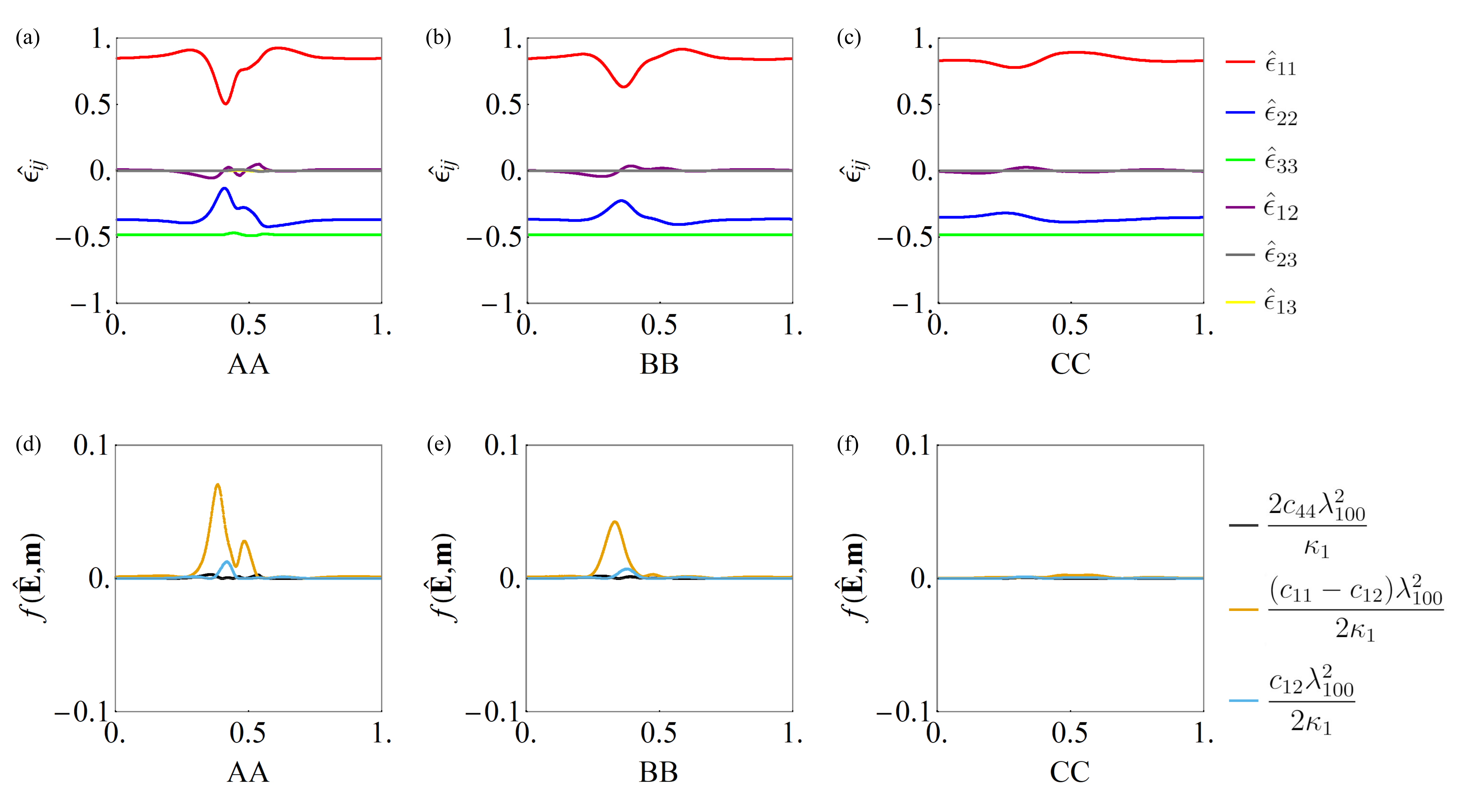}
\caption{\textcolor{black}{The strain and elastic energy distribution across the needle domain. (a-c) The strain distribution $\hat{\epsilon}_{ij}$ across lengths `AA', `BB', `CC' (as marked in Fig.~\ref{Fig:6}(d)) in the computational domain. The strain distribution from the simulation, $\hat{\epsilon}_{11} \approx 1$, $\hat{\epsilon}_{22} \approx \hat{\epsilon}_{33} \approx -0.5$, is consistent with the theoretical analysis in Fig.~\ref{Fig:6}(a). The strains at the domain walls near the spike domain vary indicating the diffuseness of the walls. The shear strains are mostly zero throughout the computational domain. (d-f) A plot of the three polynomials $f(\hat{\mathbf{E}},\mathbf{m})$  that accompany coefficients $\frac{2c_{44} \lambda_{100}^2}{\kappa_1}$, $\frac{(c_{11}-c_{12})\lambda_{100}^2}{2\kappa_1}$, and $\frac{c_{12}\lambda_{100}^2}{2\kappa_1}$ in Eq.~\ref{eq:EnergyIndicialNotation}. In line with our hypothesis the polynomials accompanying $\frac{2c_{44} \lambda_{100}^2}{\kappa_1}$ and $\frac{c_{12}\lambda_{100}^2}{2\kappa_1}$ make negligible energy contributions for magnetization reversal at the transient state of the spike domain, and the polynomial accompanying $\frac{(c_{11}-c_{12})\lambda_{100}^2}{2\kappa_1}$ serves as an energy barrier for magnetization reversal at the transient domain state. }\label{Fig:7}}
\end{figure}

The parabolic form of the locus of lowest coercivity points in Figs.~\ref{Fig:3}(a) and \ref{Fig:2}(a) can now be understood heuristically.  The terms involving the very small
nondimensional exchange constant and very large multiplier 
of the nondimensional demagnetization energy are
the only terms that involve $\nabla \bfm$.  These are expected  to compete within the diffuse domain walls present in these
simulations, but otherwise lead only to a near divergence-free
magnetization, as suggested by the
inequality of arithmetic-geometric means in the form
\beq 
\left(\sqrt{\frac{A}{\ell^2 \kappa_1}} {\rm m_{i,j}} - \sqrt{\frac{\mu_0}{6 \kappa_1}} |\nabla \zeta_{\bfm}| \delta_{ij}\right)\left(\sqrt{\frac{A}{\ell^2 \kappa_1}} {\rm m_{i,j}} - \sqrt{\frac{\mu_0}{6 \kappa_1}} |\nabla \zeta_{\bfm}| \delta_{ij} \right) \ge 0,  
\eeq
which implies, using the magnetostatic equations $\bfH_{\bfd} = - \nabla \zeta_{\bfm}$ and  $\nabla \cdot (- \nabla \zeta_{\bfm} + m_s \bfm ) =0$, that
\beq 
\frac{A}{\ell^2 \kappa_1} \rm{m_{i,j}m_{i,j}} - {\frac{\mu_0 m_s}{2 \kappa_1}} \bfH_{\bfd} \cdot \bfm \ge  2 \sqrt{\frac{A\mu_0}{6 \ell^2 \kappa_1^2}} ({\rm div}\, \bfm)|\nabla \zeta_{\bfm}|.  
\eeq

In the
geometry being considered, the heuristic argument for the
energy barrier suggests the shear strains in this geometry
play a minor role, and this is supported by typical 
measurements of the shear strains across the spike, Fig.~\ref{Fig:6}(e), taken from the simulations. This
observation, together with the fact that the magnetization
is near the easy axes $[100]$ or $[010]$ in region of the
spikes (so that $m_i m_j \approx 0, \ i \ne j$), indicates
that the constant  $c_{44} \textcolor{black}{\lambda_{100}}^2/\kappa_1$ will
play only a minor role in determining the energy barrier.  

Granted these approximations, the height of the energy barrier must then be affected mainly by the remaining dimensionless constants  
\beq
\frac{(c_{11} - c_{12}) \lambda_{100}^2}{2 \kappa_1} \quad {\rm and}
\quad \frac{c_{12} \lambda_{100}^2}{2 \kappa_1}.
\label{2consts}
\eeq
Again, from the simulations (Fig.~\ref{Fig:6}(e))
tr$\ \eps$ is quite small in comparison to the multiplier
of $(c_{11} - c_{12})\lambda_{100}^2$, and so we expect
the latter, ${(c_{11} - c_{12}) \lambda_{100}^2}/{2 \kappa_1}$,
to be the key dimensionless constant that governs the
height of the barrier. In simulations, the term containing
this constant accounted for up to 80\% of the total magnetoelastic
energy. 

\textcolor{black}{The value of this dimensionless constant giving the lowest energy barrier indicated by
the simulations is
\beq
\frac{(c_{11} - c_{12}) \lambda_{100}^2}{2 \kappa_1} = 81.0056
\label{para}
\eeq
}
As we have noticed previously \cite{renuka2021solution},
78.5\% Ni Permalloy does not exactly fall on the computed parabola given by Eq.~\ref{para}.  This may indicate that there are opportunities for lowering the coercivity of permalloy.
It is possible that favorable heat treatments of permalloy
do just that by modifying the material constants present in 
Eq.~\ref{para}.  An alternative possibility is that, while 
Eq.~\ref{para} may be the primary dimensionless constant affecting coercivity, both constants given in Eq.~\ref{2consts} may play a role, indicating that the fine tuning of elastic
moduli according to their relative roles in the two constants
Eq.~\ref{2consts} may be a route to even lower coercivity.

\section*{Discussion \label{sec:Discussion}}

At present, the conventional approach to develop soft magnets is to reduce the \textcolor{black}{magnetocrystalline anisotropy} value to zero. Consequently the search for soft magnets is concentrated around the $\kappa_1\to0$ region. Our theory-guided prediction suggests that in addition to the coercivity well at $\kappa_1\to0$ region, there exists other regions along  $\frac{(c_{11}-c_{12})\lambda_{100}^2}{2 \kappa_1} \approx 81$ at which coercivity is small. This analytical relation between material constants gives greater freedom for alloy development and increases the parameter space to discover novel soft magnets. In summary, our findings show that the magnetostriction constants, in addition to the \textcolor{black}{magnetocrystalline anisotropy} constant, play an important role in governing magnetic coercivity. This was the case in study 1 when coercivity was minimum along a parabolic relation given by   $\frac{(c_{11}-c_{12})\lambda_{100}^2}{2 \kappa_1} \approx 81$  with $\lambda_{111}=0$ in magnetic alloys. In study 2, we identified a 3D surface topology on which the coercivity is small. Below, we discuss some limitations of our findings and then present its potential impact to the magnetic alloy development program.

Two features of this work limit the conclusions we can draw about the fundamental relationship between magnetic material constants. First, we compute coercivities assuming a simple defect structure (i.e., two spike-domains formed around a non-magnetic inclusion) in an oblate ellipsoid. While defect geometry has been shown to have surprisingly small effect on the coercivity values \cite{balakrishna2021tool}, it is not known whether and how different defect geometries, \textcolor{black}{and crystallographic texture of the magnetic alloy} would affect our predictions of the parabolic relation for minimum coercivity. Second, the presence of mechanical loads, such as residual strains or boundary loads, could affect the delicate balance between the magnetic material constants. The idealized stress-free conditions in our model is subject to the shortcomings associated with the presence of microstructural inhomogeneities and surface conditions in bulk materials. Whether accounting for these inhomogeneities would yield comparable results in experiments is an open question. With these reservations in mind we next discuss the impact of our findings to the alloy development program.

A key feature of our work is that we demonstrate that the magnetostriction constants play an important role in governing magnetic coercivity. These findings contrast with prior research in which the \textcolor{black}{magnetocrystalline anisotropy} constant has been regarded as the $\mathit{only}$ material parameter that governs magnetic hysteresis, and the contribution from magnetostriction constants have been largely neglected. Consequently, the commonly accepted norm in the literature is that magnetic alloys with small \textcolor{black}{magnetocrystalline anisotropy} constant have small coercivity, and magnetic alloys with large \textcolor{black}{magnetocrystalline anisotropy} constant have large coercivity. However, by accounting for both \textcolor{black}{magnetocrystalline anisotropy} and magnetostriction constants, we show that magnetic alloys despite their large $\kappa_1$ values have small coercivities at specific combinations of the magnetostriction constants.

Another significant feature of our work is that we identify a relationship   $\frac{(c_{11}-c_{12})\lambda_{100}^2}{2 \kappa_1} \approx 81$  between magnetic material constants at which the coercivity is small. This generic formula serves as a theoretical guide to the alloy development program, by suggesting alternative combinations of material constants---beyond $\kappa_1=0$---to develop soft magnets. \textcolor{black}{While this relation is based on $\langle 100 \rangle$ easy axes and might differ for alloys with different easy axes, our work is intended to guide the search for soft magnetism in ordinary ferromagnets with large $m_s$. In future work, we intend to investigate other easy axes, elastic stiffness constants, to further lower coercivity in magnetic alloys \cite{ahani2021}.} With the recent advances in atomic-scale engineering the compositions of magnetic alloys can be tuned atom-by-atom \cite{ouazi2012atomic, kablov2015bifurcation}. For these experimental approaches, our prediction of the material constant formula could serve as a guiding principle, to engineer magnetic alloy compositions to small hysteresis. Overall, the fundamental relationship between material constants provides initial steps to experimentalists to discover soft magnets with high \textcolor{black}{magnetocrystalline anisotropy} constants.

\vspace{2mm}
\noindent In conclusion, the present findings contribute to a more nuanced understanding of how material constants, such as \textcolor{black}{magnetocrystalline anisotropy} and magnetostriction constants, affect magnetic hysteresis. Specifically, magnetoelastic interactions have been regarded to play a negligible role in lowering magnetic coercivity. Given the current findings, we quantitatively demonstrate that the delicate balance between \textcolor{black}{magnetocrystalline anisotropy}, magnetostriction constants and the spike domain microstructure (localized disturbance) is necessary to lower magnetic coercivity. 
{\textcolor{black}{We propose a mathematical relationship between material constants   $\frac{(c_{11}-c_{12})\lambda_{100}^2}{2 \kappa_1} \approx 81$  at} which minimum coercivity can be achieved in a material with $(100)$ easy axes. More generally, our findings serve as a theoretical guide to discover novel combinations of material constants that lower coercivity in magnetic alloys.

\section*{Methods \label{sec:Methods}}

\subsection*{Micromagnetics}
\textcolor{black}{In our coercivity tool, we use micromagnetics theory
that describes the total free energy as a function of magnetization
$\mathbf{m}$, strain $\mathbf{E}$, and magnetostatic field $\mathbf{H_{d}}$}:

\textcolor{black}{
\begin{align}
\mathcal{\psi} & =\int_{\Omega}\{\nabla\mathbf{m}\cdot\mathrm{A\nabla\mathbf{m}+\kappa_{1}(\mathrm{m_{1}^{2}m_{2}^{2}}+\mathrm{m_{2}^{2}m_{3}^{2}}+\mathrm{m_{3}^{2}m_{1}^{2}})}+\frac{1}{2}[\mathbf{E}-\mathbf{E_{0}\mathrm{(\mathbf{m})}}]\cdot\mathbb{C}[\mathbf{E}-\mathbf{E_{0}\mathrm{(\mathbf{m})}}]-\mu_{0}m_s\mathbf{H_{\mathrm{ext}}\cdot m}\}\mathrm{d\mathbf{x}}\nonumber \\
 & +\int_{\mathbb{R}^{3}}\frac{\mu_{0}}{2}\left|\mathbf{H_{d}}\right|^{2}\mathrm{d\mathbf{x}}.  \label{eq:FreeEnergy}
\end{align}
}

\textcolor{black}{The form of Eq. \ref{eq:FreeEnergy} is identical to that used in
our previous work, in which we detail the meaning of the specific terms, material constants, and normalizations \cite{balakrishna2021tool}. For the present work, we note that
the exchange energy $\nabla\mathbf{m}\cdot\mathrm{A\nabla\mathbf{m}}$
penalizes spatial gradients of magnetization. \textcolor{black}{More generally, the anisotropy energy for cubic alloys would have contributions from higher order energy terms (e.g., $\kappa_2(\mathrm{m_1^2m_2^2m_3^2})$), and these additional anisotropy coefficients are likely in general to contribute to our coercivity calculations. Although including these higher order anisotropy coefficients, e.g. $\kappa_2$, could affect the easy axes of magnetization of the material. For example at $\kappa_1 \approx \frac{-\kappa_2}{2}$ other crystallographic directions, such as $[111], [110]$, and a family of irrational directions, would have similarly small magnetocrystalline anisotropy energy. This warrants a systematic investigation in a future study, especially with $\kappa_1, \kappa_2, \kappa_3$ near actual measured values. However, we do not think that these higher order terms would significantly affect our coercivity calculations for the following reasons: First, in our computations, we model an oblate ellipsoid (pancake-shaped) that supports an in-plane magnetization. This ellipsoid geometry and the defect shape penalizes out-of-plane magnetization and thus $\mathrm{m_3} \approx 0$ in our calculations of the magnetic hysteresis. Consequently, the energy contribution from the higher order anisotropy term, $\kappa_2(\mathrm{m_1^2m_2^2m_3^2})$, is negligible in our computations. Second, in the nondimensional form of the free energy (see Eq. (5)), the energy contribution from the higher-order anisotropy term would scale as $\frac{\kappa_2}{\kappa_1}(\mathrm{m_1^2m_2^2m_3^2})$ with $|\mathbf{m}| = 1$. This  sixth-order energy term is expected not to change the coercivity calculations significantly. We propose to investigate the precise role of the higher-order anisotropy terms in a future study, however, as a first step we investigate coercivity as a function of magnetocrystalline anisotropy $\kappa_1$ and magnetostriction constant $\lambda_{100}$.} The elastic energy $\frac{1}{2}[\mathbf{E}-\mathbf{E_{0}\mathrm{(\mathbf{m})}}]\cdot\mathbb{C}[\mathbf{E}-\mathbf{E_{0}\mathrm{(\mathbf{m})}}]$
penalizes mechanical deformation away from the preferred strains,
and the external energy, $\mu_{0}\mathbf{H_{\mathrm{ext}}\cdot m}$
accounts for the mutual interaction between magnetization moment and
the applied field. Finally, the magnetostatic energy $\frac{\mu_{0}}{2}\left|\mathbf{H_{d}}\right|^{2}$
computed in all of space $\mathbb{R}^{3}$ penalizes the stray fields
generated by the magnetic body in its surroundings. }

\textcolor{black}{We compute the evolution of the magnetization using an energy minimization technique, the generalized Landau-Lifshitz-Ginzburg
equation, see Fig. \ref{Fig:4}:}

\begin{equation}
\frac{\partial\mathbf{m}}{\partial \tau}=-\mathbf{m}\times\mathcal{H}-\alpha \mathbf{m}\times(\mathbf{m}\times\mathcal{H}).\label{eq:LLG}
\end{equation}

\textcolor{black}{Here, $\mathcal{H}=-\frac{1}{\mu_0m_s^2}\frac{\delta\Psi}{\delta\mathbf{m}}$
is the effective field, $\tau=\gamma m_s t$ is teh dimensionless time step, $\gamma$ is the gyromagnetic ratio, and $\alpha$
is the damping constant. We numerically solve Eq. \ref{eq:LLG} using the Gauss
Siedel projection method  \cite{wang2001gauss}, and identify equilibrium states when the magnetization evolution converges, $\left|\mathbf{m^{\mathrm{n+1}}}-\mathrm{\mathbf{m}^{n}}\right|<10^{-9}$.
At each iteration we compute the magnetostatic field $\mathbf{H_{d}}=-\nabla\zeta_{\mathbf{m}}$
and the strain  $\mathbf{E}$ by solving their respective equilibrium
equations:}

\textcolor{black}{
\begin{align}
\nabla\cdot(\mathbf{\mathbf{-\nabla\zeta_{\mathbf{m}}}+\mathit{m_s}m}) & =0\qquad\mathrm{on}\thinspace\mathbb{R}^{3}\label{eq:magnetostaticeq}\\
\nabla\cdot\mathbb{C}(\mathbf{E-E}_{0}) & =0\qquad\mathrm{on}\thinspace\mathcal E.\label{eq:mechanicaleq}
\end{align}
}

The magnetostatic equilibrium condition arises from
the Maxwell equations, namely $\nabla\times\mathbf{H_{d}}=0\to\mathbf{H_{d}}=-\nabla\zeta_{\mathbf{m}}$
and $\nabla\cdot\mathbf{B}=\nabla\cdot(\mathbf{H_{d}+\mathit{m_s}m})=0$.

In our calculations, we model a finite sized domain $\Omega$ centered
around a non-magnetic defect $\Omega_{d}$, see inset images in Fig. \ref{Fig:4}. This domain is several times smaller
than the actual size of the ellipsoid $\mathcal{E},$ see Fig. \ref{Fig:1}. We define the total demagnetization field as a sum of the local $\mathbf{\widetilde{H}}(\mathbf{x})$ and non-local contributions $\mathbf{\overline{H}}$. The local contribution $\mathbf{\widetilde{H}}(\mathbf{x})$ varies spatially and accounts for the magnetostatic fields generated from
defects and other imperfections inside the body. We calculate this local contribution by solving $\nabla\cdot(\mathbf{\widetilde{H}+\mathit{m_s}\widetilde{m}})=0$ on $\Omega$. The non-local contribution is computed as $\mathbf{\overline{H}}=-\mathbf{N\mathit{m_s}\overline{m}}$. \textcolor{black}{Here, $\mathbf{N}$ is the demagnetization factor matrix that is a
tabulated geometric property of the ellipsoid. We note, from the tabulated values in Ref. \cite{osborn1945demagnetizing}, the demagnetization factors for an oblate ellipsoid (pancake-shaped) are $\mathrm{N_{11}=N_{22}=0, N_{33}=1}$.} The $\mathbf{\overline{m}}$ is the constant magnetization which is defined such that $\int_{\Omega}\mathbf{m}(\mathbf{x})\mathrm{d}\mathbf{x}=0$. This decomposition simplifies our
computational complexity, because we now model a local domain $\Omega$ that is much smaller than modeling a domain in  $\mathbb{R}^{3}$, and yet account for the demagnetization contributions from both the body geometry and the local defects. This decomposition is justified in the appendix of our previous paper \cite{balakrishna2021tool}. Both the magnetostatic and mechanical
equilibrium conditions in Eqs. \ref{eq:magnetostaticeq}--\ref{eq:mechanicaleq} are solved in Fourier space, see Refs.~\cite{balakrishna2021tool, zhang2005phase} for further details.

\subsection*{Numerical calculations}
\textcolor{black}{In the present work, we calibrate our micromagnetics
model for the FeNi alloy with the following material constants: $A=10^{-11}\mathrm{Jm^{-1}}, m_s=10^{6}\mathrm{Am^{-2}}, \mu_0=1.3 \times 10^{-6}\mathrm{NA^{-2}} , c_{11}=240.8\mathrm{GPa}, c_{12}=89.2\mathrm{GPa}, c_{44}=75.8\mathrm{GPa}$. The values of $\kappa_1,\lambda_{100},\lambda_{111}$ are systematically varied as detailed in the results section. Here, note that magnetic alloys with positive and negative \textcolor{black}{magnetocrystalline anisotropy} constants,
have their easy axes along $\{100\}$ and $\{111\}$ crystallographic directions, respectively. To accommodate this change of easy axes, we transform the energy potential from a cubic basis, and further details of this transformation are described in the appendix of \cite{balakrishna2021tool}}.

We compute the coercivity of magnetic ellipsoids as a function of the \textcolor{black}{magnetocrystalline anisotropy} and magnetostriction material constants. In each computation, we model a domain of size $64\times64\times24$ containing a defect with $16\times16\times6$ grid points. We choose a grid size such that the domain walls span across 3-4 unit cells. We initialize the computational domain with a uniform magnetization $\mathrm{\mathbf{m}=\mathrm{m_{1}}}$, and force the magnetization inside the defect to be zero throughout the computation, $\mathrm{\left|\mathbf{m}\right|=\mathrm{0}}$, see Fig. \ref{Fig:4}. We apply a large external field along the easy axes, $\mathbf{H}_{\mathrm{ext}}>>0$, and decrease it gradually in steps of $\delta \mathbf{H} = 0.25\mathrm{e_1} \mathrm{Oe}$. As we decrease the applied field a spike domain forms around the defect, see Fig. \ref{Fig:1}. This spike domain grows in size as the applied field is further lowered until a critical field value---known as the coercivity $\mathrm{H_{\mathrm{c}}}$---at which the magnetization vector reverses. We use this approach to predict the coercivity of the magnetic alloys at each combination of the \textcolor{black}{magnetocrystalline anisotropy} and magnetostriction constants.

\textcolor{black}{Specifically, in Study 1 and Study 2 we carry out a total of, $n=2,163$ and $n=605$, computations respectively. In these computations, we systematically vary the material constants in the parameter space range of $0 \leq \kappa_1 \leq 2000\mathrm{Jm^{-3}}$, $-2000\mu\epsilon \leq \lambda_{100} \leq 2000\mu\epsilon$, and $0 \leq \lambda_{111} \leq 600\mu\epsilon$, respectively. Our investigation shows that the minimum coercivity is attained for a parabolic relation $\kappa_1 \propto \lambda_{100}^2$, and a total of over 2,500 computations are necessary to confirm this relationship.}

\vspace{2mm}
\noindent {\bf Data Availability}. The authors declare that the data supporting the findings of this study are available within the paper and its supplementary information files. Furthermore, additional data that support the findings of this study are available from the corresponding author upon request.

\vspace{2mm}
\noindent {\bf Acknowledgment}. The authors acknowledge the Center for Advanced Research Computing at the University of Southern California and the Minnesota Supercomputing Institute at the University of Minnesota for providing resources that contributed to the research results reported within this paper. The authors would like to thank Anjanroop Singh (University of Minnesota) for help in checking some of the calculations. A.R.B acknowledges the support of a Provost Assistant Professor Fellowship, Gabilan WiSE fellowship, and USC's start-up funds. R.D.J acknowledges the support of a Vannevar Bush Faculty Fellowship. The authors thank NSF (DMREF-1629026) and ONR (N00014-18-1-2766) for partial support of this work.

\vspace{2mm}
\noindent {\bf Author contribution}. ARB and RDJ conceptualized the project, designed the methodology, and procured funding. ARB worked on model development, theoretical analysis, and visualization of data. RDJ worked on the theoretical calculations. Both authors were involved with the writing of the paper. 

\vspace{2mm}
\noindent {\bf Competing Interests}. The authors declare that there are no competing interests.

\printbibliography

@article{fert2008nobel,
  title={Nobel Lecture: Origin, development, and future of spintronics},
  author={Fert, Albert},
  journal={Reviews of modern physics},
  volume={80},
  number={4},
  pages={1517},
  year={2008},
  publisher={APS}
}

@article{bozorth1953permalloy,
  title={The permalloy problem},
  author={Bozorth, RM},
  journal={Reviews of Modern Physics},
  volume={25},
  number={1},
  pages={42},
  year={1953},
  publisher={APS}
}

@article{takahashi1987magnetocrystalline,
  title={Magnetocrystalline anisotropy and magnetostriction of Fe-Si-Al (Sendust) single crystals},
  author={Takahashi, M and Nishimaki, S and Wakiyama, T},
  journal={Journal of magnetism and magnetic materials},
  volume={66},
  number={1},
  pages={55--62},
  year={1987},
  publisher={Elsevier}
}

@book{brown1963micromagnetics,
  title={Micromagnetics},
  author={Brown, William Fuller},
  number={18},
  year={1963},
  publisher={interscience publishers}
}

@article{silveyra2018soft,
  title={Soft magnetic materials for a sustainable and electrified world},
  author={Silveyra, Josefina M and Ferrara, Enzo and Huber, Dale L and Monson, Todd C},
  journal={Science},
  volume={362},
  number={6413},
  year={2018},
  publisher={American Association for the Advancement of Science}
}

@book{brown1962magnetostatic,
  title={Magnetostatic principles in ferromagnetism},
  author={Brown, William Fuller},
  volume={1},
  year={1962},
  publisher={North-Holland Publishing Company}
}

@article{aharoni1958magnetization,
  title={Magnetization curve of the infinite cylinder},
  author={Aharoni, A and Shtrikman, S},
  journal={Physical Review},
  volume={109},
  number={5},
  pages={1522},
  year={1958},
  publisher={APS}
}

@article{wang2001gauss,
  title={A Gauss--Seidel projection method for micromagnetics simulations},
  author={Wang, Xiao-Ping and Garc{\i}a-Cervera, Carlos J and Weinan, E},
  journal={Journal of Computational Physics},
  volume={171},
  number={1},
  pages={357--372},
  year={2001},
  publisher={Elsevier}
}

@article{teter1990magnetostriction,
  title={Magnetostriction and hysteresis for Mn substitutions in (Tb$_x$Dy$_{1-x}$)(Mn$_y$Fe$_{1-y}$)$_1.95$},
  author={Teter, JP and Clark, AE and Wun-Fogle, M and McMasters, OD},
  journal={IEEE transactions on magnetics},
  volume={26},
  number={5},
  pages={1748--1750},
  year={1990},
  publisher={IEEE}
}

@article{clark1988magnetostriction,
  title={Magnetostriction ‘‘jumps’’ in twinned Tb$_{0.3}$Dy$_{0.7}$Fe$_{1.9}$},
  author={Clark, AE and Teter, JP and McMasters, OD},
  journal={Journal of applied physics},
  volume={63},
  number={8},
  pages={3910--3912},
  year={1988},
  publisher={American Institute of Physics}
}

@article{atulasimha2011review,
  title={A review of magnetostrictive iron--gallium alloys},
  author={Atulasimha, Jayasimha and Flatau, Alison B},
  journal={Smart Materials and Structures},
  volume={20},
  number={4},
  pages={043001},
  year={2011},
  publisher={IOP Publishing}
}

@article{clark2002magnetostrictive,
  title={Magnetostrictive properties of Galfenol alloys under compressive stress},
  author={Clark, Arthur E and Wun-Fogle, Marilyn and Restorff, James B and Lograsso, Thomas A},
  journal={Materials transactions},
  volume={43},
  number={5},
  pages={881--886},
  year={2002},
  publisher={The Japan Institute of Metals and Materials}
}

@article{yang2010large,
  title={Large magnetostriction from morphotropic phase boundary in ferromagnets},
  author={Yang, Sen and Bao, Huixin and Zhou, Chao and Wang, Yu and Ren, Xiaobing and Matsushita, Yoshitaka and Katsuya, Yoshio and Tanaka, Masahiko and Kobayashi, Keisuke and Song, Xiaoping and others},
  journal={Physical review letters},
  volume={104},
  number={19},
  pages={197201},
  year={2010},
  publisher={APS}
}

@article{bergstrom2013morphotropic,
  title={Morphotropic Phase Boundaries in Ferromagnets: Tb$_{1-x}$Dy$_x$Fe$_2$ Alloys},
  author={Bergstrom Jr, Richard and Wuttig, Manfred and Cullen, James and Zavalij, Peter and Briber, Robert and Dennis, Cindi and Garlea, V Ovidiu and Laver, Mark},
  journal={Physical review letters},
  volume={111},
  number={1},
  pages={017203},
  year={2013},
  publisher={APS}
}

@article{hu2021room,
  title={Room-temperature ultrasensitive magnetoelastic responses near the magnetic-ordering tricritical region},
  author={Hu, Cheng-Chao and Zhang, Zhao and Cai, Ting-Tao and Xu, Yu-Xin and Hao, Ji-Gong and Shi, Yang-Guang and Yang, Tian-Nan and Chen, Long-Qing},
  journal={Journal of Applied Physics},
  volume={130},
  number={6},
  pages={063901},
  year={2021},
  publisher={AIP Publishing LLC}
}

@article{zhang2005phase,
  title={Phase-field microelasticity theory and micromagnetic simulations of domain structures in giant magnetostrictive materials},
  author={Zhang, JX and Chen, LQ},
  journal={Acta Materialia},
  volume={53},
  number={9},
  pages={2845--2855},
  year={2005},
  publisher={Elsevier}
}

@article{wang2013real,
  title={A real-space phase field model for the domain evolution of ferromagnetic materials},
  author={Wang, Jie and Zhang, Jianwei},
  journal={International Journal of Solids and Structures},
  volume={50},
  number={22-23},
  pages={3597--3609},
  year={2013},
  publisher={Elsevier}
}

@article{knupfer2013nucleation,
  title={Nucleation barriers for the cubic-to-tetragonal phase transformation},
  author={Kn{\"u}pfer, Hans and Kohn, Robert V and Otto, Felix},
  journal={Communications on pure and applied mathematics},
  volume={66},
  number={6},
  pages={867--904},
  year={2013},
  publisher={Wiley Online Library}
}

@article{zhang2009energy,
  title={Energy barriers and hysteresis in martensitic phase transformations},
  author={Zhang, Zhiyong and James, Richard D and M{\"u}ller, Stefan},
  journal={Acta Materialia},
  volume={57},
  number={15},
  pages={4332--4352},
  year={2009},
  publisher={Elsevier}
}

@phdthesis{pilet2006relation,
  title={The relation between magnetic hysteresis and the micromagnetic state explored by quantitative magnetic force microscopy},
  author={Pilet, Nicolas},
  year={2006},
  school={University of Basel}
}

@article{schafer2020tomography,
  title={Tomography of basic magnetic domain patterns in ironlike bulk material},
  author={Sch{\"a}fer, R and Schinnerling, S},
  journal={Physical Review B},
  volume={101},
  number={21},
  pages={214430},
  year={2020},
  publisher={APS}
}

@article{soldatov2020inverted,
  title={Inverted Hysteresis, Magnetic Domains, and Hysterons},
  author={Soldatov, Ivan and Andrei, Petru and Schaefer, Rudolf},
  journal={IEEE Magnetics Letters},
  volume={11},
  pages={1--5},
  year={2020},
  publisher={IEEE}
}

@book{hubert2008magnetic,
  title={Magnetic domains: the analysis of magnetic microstructures},
  author={Hubert, Alex and Sch{\"a}fer, Rudolf},
  year={2008},
  publisher={Springer Science \& Business Media}
}

@article{thomas2020nanocrystallites,
  title={Nanocrystallites via Direct Melt Spinning of Fe$_{77}$Ni$_{5.5}$Co$_{5.5}$Zr$_7$B$_4$Cu for Enhanced Magnetic Softness},
  author={Thomas, Som V and Willard, Matthew A and Martone, Anthony and Heben, Michael J and Solomon, Virgil and Welton, Aaron and Boolchand, Punit and Ewing, Rodney C and Wang, Chenxu and Bud'ko, Sergey L and others},
  journal={physica status solidi (a)},
  volume={217},
  number={8},
  pages={1900680},
  year={2020},
  publisher={Wiley Online Library}
}

@article{mchenry1999amorphous,
  title={Amorphous and nanocrystalline materials for applications as soft magnets},
  author={McHenry, Michael E and Willard, Matthew A and Laughlin, David E},
  journal={Progress in materials Science},
  volume={44},
  number={4},
  pages={291--433},
  year={1999},
  publisher={Elsevier}
}

@article{osborn1945demagnetizing,
  title={Demagnetizing factors of the general ellipsoid},
  author={Osborn, JA},
  journal={Physical review},
  volume={67},
  number={11-12},
  pages={351},
  year={1945},
  publisher={APS}
}

@article{balakrishna2021tool,
  title={A tool to predict coercivity in magnetic materials},
  author={Balakrishna, Ananya Renuka and James, Richard D},
  journal={Acta Materialia},
  volume={208},
  pages={116697},
  year={2021},
  publisher={Elsevier}
}

@article{renuka2021solution,
  title={A solution to the permalloy problem—A micromagnetic analysis with magnetostriction},
  author={Renuka Balakrishna, Ananya and James, Richard D},
  journal={Applied Physics Letters},
  volume={118},
  number={21},
  pages={212404},
  year={2021},
  publisher={AIP Publishing LLC}
}

@book{brown1966magnetoelastic,
  title={Magnetoelastic interactions},
  author={Brown, William Fuller},
  volume={9},
  year={1966},
  publisher={Springer}
}

@article{cui2006combinatorial,
  title={Combinatorial search of thermoelastic shape-memory alloys with extremely small hysteresis width},
  author={Cui, Jun and Chu, Yong S and Famodu, Olugbenga O and Furuya, Yasubumi and Hattrick-Simpers, Jae and James, Richard D and Ludwig, Alfred and Thienhaus, Sigurd and Wuttig, Manfred and Zhang, Zhiyong and others},
  journal={Nature materials},
  volume={5},
  number={4},
  pages={286--290},
  year={2006},
  publisher={Nature Publishing Group}
}

@article{tickle1999magnetic,
  title={Magnetic and magnetomechanical properties of Ni2MnGa},
  author={Tickle, R and James, Richard D},
  journal={Journal of Magnetism and Magnetic Materials},
  volume={195},
  number={3},
  pages={627--638},
  year={1999},
  publisher={Elsevier}
}

@article{zwicknagl2014microstructures,
  title={Microstructures in low-hysteresis shape memory alloys: scaling regimes and optimal needle shapes},
  author={Zwicknagl, Barbara},
  journal={Archive for Rational Mechanics and Analysis},
  volume={213},
  number={2},
  pages={355--421},
  year={2014},
  publisher={Springer}
}

@article{james2000martensitic,
  title={Martensitic transformations and shape-memory materials},
  author={James, Richard D and Hane, Kevin F},
  journal={Acta materialia},
  volume={48},
  number={1},
  pages={197--222},
  year={2000},
  publisher={Elsevier}
}

@article{otto2010domain,
  title={Domain branching in uniaxial ferromagnets: asymptotic behavior of the energy},
  author={Otto, Felix and Viehmann, Thomas},
  journal={Calculus of variations and partial differential equations},
  volume={38},
  number={1},
  pages={135--181},
  year={2010},
  publisher={Springer}
}

@article{otto2010concertina,
  title={The concertina pattern},
  author={Otto, Felix and Steiner, Jutta},
  journal={Calculus of variations and partial differential equations},
  volume={39},
  number={1},
  pages={139--181},
  year={2010},
  publisher={Springer}
}

@article{doring2014reduced,
  title={A reduced model for domain walls in soft ferromagnetic films at the cross-over from symmetric to asymmetric wall types},
  author={D{\"o}ring, Lukas and Ignat, Radu and Otto, Felix},
  journal={Journal of the European Mathematical Society},
  volume={16},
  number={7},
  pages={1377--1422},
  year={2014}
}

@article{cinti2016interpolation,
  title={Interpolation inequalities in pattern formation},
  author={Cinti, Eleonora and Otto, Felix},
  journal={Journal of Functional Analysis},
  volume={271},
  number={11},
  pages={3348--3392},
  year={2016},
  publisher={Elsevier}
}

@article{ahani2021,
  title={Magnetoelastic interactions reduce hysteresis in soft magnets},
  author={Ahani, Negar and Daware, Ashwin and Balakrishna, Ananya Renuka},
  journal={Manuscript in preparation},
  volume={},
  number={},
  pages={},
  year={2021},
  publisher={}
}

@article{ouazi2012atomic,
  title={Atomic-scale engineering of magnetic anisotropy of nanostructures through interfaces and interlines},
  author={Ouazi, S and Vlaic, S and Rusponi, S and Moulas, G and Buluschek, P and Halleux, K and Bornemann, S and Mankovsky, S and Min{\'a}r, J and Staunton, Julie B and others},
  journal={Nature communications},
  volume={3},
  number={1},
  pages={1--9},
  year={2012},
  publisher={Nature Publishing Group}
}

@article{kablov2015bifurcation,
  title={Bifurcation of magnetic anisotropy caused by small addition of Sm in (Nd$_{1-x}$Sm$_x$Dy)(FeCo)B magnetic alloy},
  author={Kablov, EN and Ospennikova, OG and Kablov, DE and Piskorskii, VP and Kunitsyna, EI and Dmitriev, AI and Valeev, RA and Korolev, DV and Rezchikova, II and Talantsev, AD and others},
  journal={Journal of Applied Physics},
  volume={117},
  number={24},
  pages={243903},
  year={2015},
  publisher={AIP Publishing LLC}
}

\end{document}


\maketitle

\section*{Supplementary Methods}
\textcolor{black}{In our calculations, we compute coercivity for an oblate ellipsoid $\calE$ (pancake shaped) under a suitably oriented applied field. We apply the external magnetic field along the major axis of the ellipsoid that is uniformly magnetized $\mathbf{m}$ except at the proximity of the defects. We account for the effect of the critically important, far-away poles on the boundary of the macroscopic body, while treating their influence on the high-resolution computational domain near the defect, using the ellipsoid and reciprocal theorems. In our numerical scheme, we use these theorems, to decompose the magnetization into two: a constant magnetization $\overline{\mathbf{m}}$, and a spatially varying magnetization $\widetilde{\mathbf{m}}$. The latter magnetization is localized in the vicinity of the defect and decays away from it.}
\vspace{2mm}

\begin{figure}[h]
    \centering
    \includegraphics[width=\textwidth]{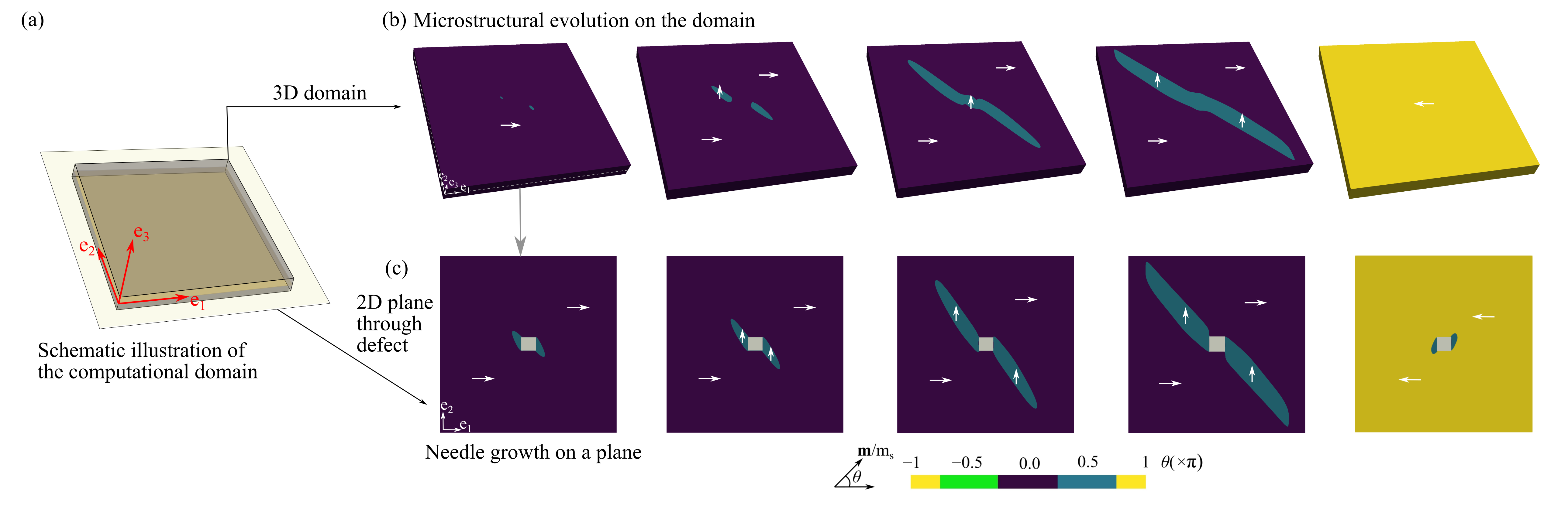}
    \caption{A representative micromagnetic simulation. (a) A schematic illustration of the 3D computational domain used in our numerical calculations. (b) The 3D microstructural evolution of the spike-domain under an applied magnetic field. (c) The spike-domain growth on a $\mathbf{e}_1-\mathbf{e}_2$ plane through the 3D computational domain.}
    \label{fig:3Dmicrostructure-supplement}
\end{figure}

\textcolor{black}{The ellipsoid geometry of the magnetic body is essential as a way of describing the poles on the boundary of the macroscale body without having to simulate the external fields arising due to these poles. As a consequence of this ellipsoid theorem, we model a 3D computational domain $\Omega >> \calE$, with a nonmagnetic inclusion at its geometric centre. The size of this computational domain is chosen such that $\widetilde{\mathbf{m}} \to \overline{\mathbf{m}}$ on the surface of the computational domain. Further details on how the magnetization and the derivative fields are computed are described in Ref. \cite{balakrishna2021tool}, and for our purposes we note that the equilibrium conditions and evolution equations Eqs.~11--13 are solved using the Gauss-Siedel Projection method in Fourier space \cite{wang2001gauss, zhang2005phase, balakrishna2021tool}.}

\textcolor{black}{Supplementary Figure~\ref{fig:3Dmicrostructure-supplement} shows a representative calculation with the 3D evolution of the spike domain microstructure during magnetization reversal. Please note that we model an oblate ellipsoid in our computations. This ellipsoid geometry of the magnetic body forces an in-plane magnetization, and consequently the domain evolution in our computations are primarily observed on the $\mathbf{e}_1-\mathbf{e}_2$ plane with negligible changes in the third dimension.}

\section*{Supplementary Discussion}

\begin{figure}
    \centering
    \includegraphics[width=0.8\textwidth]{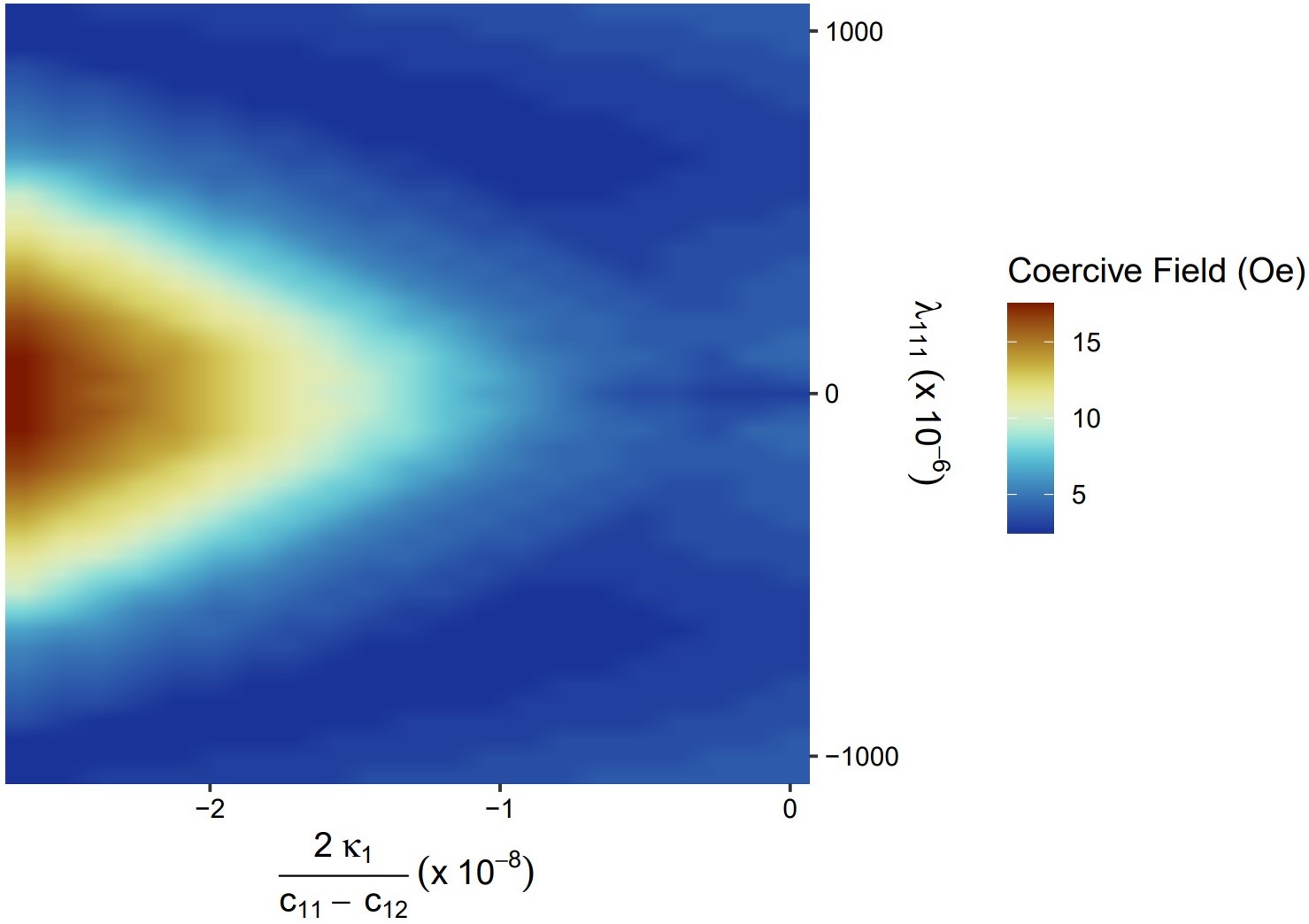}
    \caption{A heat map of the coercivity values as a function of the anisotropy constant $\kappa_{1}$ and magnetostriction constant $\lambda_{111}$. We set the magnetostriction constant along the $<100>$ crystallographic directions $\lambda_{100}$ to be zero in these computations $n=903$. We find that minimum coercivity is achieved for suitable combinations of \textcolor{black}{magnetocrystalline anisotropy} and magnetostriction constants, $\kappa_1 \propto \lambda_{111}^2$, for magnetic alloys with $\kappa_1 < 0$.}
    \label{fig:K1_L111_supplement}
\end{figure}

In our micromagnetic calculations, we primarily investigated coercivities of magnetic alloys with easy axes along $<100>$ crystallographic directions (i.e., $\kappa_1 > 0$). In this supplement we investigate whether suitable magnetostrictions can lower coercivities of magnetic alloys with easy axes along $<111>$ crystallographic directions (i.e., $\kappa_1 < 0 $). We model magnetic ellipsoids with with crystallographic directions $[111]$ and $[\overline{1}10]$ in-plane, and apply an external field along the [111] direction. We transform the coordinate basis of the free energy function to suitable orientations (as described in the Methods section) and further details of this transformation are given in Ref.~\cite{balakrishna2021tool}. We compute coercivity by systematically varying the \textcolor{black}{magnetocrystalline anisotropy} and the magnetostriction constants, in the ranges $ -2000 \leq \kappa_1 \leq 0\mathrm{Jm^{-3}}$,= and $ -1050\mu\epsilon \leq \lambda_{111} \leq 1050\mu\epsilon $, respectively. Further details on the method, and specific values of the material constants used in our model are described in the methods section. Overall, our computations from over 903 independent calculations show that the lowest coercivity is attained when  $\kappa_1 \propto \lambda_{111}^2$, see Supplementary Figure~\ref{fig:K1_L111_supplement}. This is consistent with our theoretical and analytical interpretation on the role of magnetostriction constant in lowering coercivity for magnetic alloys with $\kappa_1 > 0$.

\printbibliography